%% file: main.tex
\documentclass[%
 reprint,
superscriptaddress,
 amsmath,amssymb,
 aps,
prb,
floatfix,
]{revtex4-2}

\usepackage{enumitem}% costume listings
\usepackage{graphicx}% Include figure files
\usepackage{dcolumn}% Align table columns on decimal point
\usepackage{bm}% bold math
\usepackage{xr-hyper}
\usepackage{xr}

\usepackage[unicode=true,pdfusetitle,
 bookmarks=true,bookmarksnumbered=false,bookmarksopen=false,
 breaklinks=true,pdfborder={0 0 0},pdfborderstyle={},backref=false,colorlinks=true,allcolors=blue]{hyperref}% add hypertext capabilities
\usepackage{lipsum}
\usepackage{xfrac}
\usepackage{mathtools}
\usepackage{braket}
\usepackage{enumitem}
\usepackage{mathdots}
\usepackage{bbm}
\usepackage{adjustbox}
\usepackage{appendix}

\usepackage{lmodern}
\usepackage[utf8]{inputenc}
\usepackage{float}
\usepackage{units}
\usepackage{amssymb}
\usepackage{cancel}
\usepackage{graphicx}
\usepackage{setspace}
\usepackage{braket}
\usepackage{color}
\usepackage[table]{xcolor}
{\tiny{}}
\global\long\def\bra#1{\Bra{#1}}%
{\tiny{}}
\global\long\def\ket#1{\Ket{#1}}%
{\tiny{}}
{\tiny{}}
{\tiny{}}
{\tiny{}}
{\tiny{}}
{\tiny{}}
{\tiny\par}

{\tiny{}}
{\tiny{}}
{\tiny{}}
{\tiny{}}
{\tiny{}}
{\tiny{}}
{\tiny{}}
{\tiny{}}
{\tiny{}}
{\tiny{}}
{\tiny\par}

\global\long\def\unit{\mathbf{1}}%
\usepackage{xcolor,cancel}

\begin{document}

\preprint{APS/123-QED}

\title{Defects in Graphene : A Topological Description }
%- A Landau theory for topological phase transitions }% Force line breaks with \\

\author{Amit Goft}
\affiliation{%
Department of Physics, Technion -- Israel Institute of Technology, Haifa 3200003, Israel
}%
\author{Yuval Abulafia}
\affiliation{%
Department of Physics, Technion -- Israel Institute of Technology, Haifa 3200003, Israel
}%
\author{Nadav Orion}
\affiliation{%
Department of Physics, Technion -- Israel Institute of Technology, Haifa 3200003, Israel
}%
\author{Claude L.~Schochet}
\affiliation{%
Department of Mathematics, Technion – Israel Institute of Technology, Haifa 3200003, Israel
}%
%\address{Department of Mathematics,
    %Technion,
   % Haifa 3200003, Israel}
\author{Eric Akkermans}%
% \email{eric@physics.technion.ac.il}
\affiliation{%
Department of Physics, Technion -- Israel Institute of Technology, Haifa 3200003, Israel
}%

\date{\today}% It is always \today, today,

\begin{abstract}
Specific types of spatial defects or potentials can turn monolayer graphene into a topological material. These topological defects are classified by a spatial dimension $D$ and they are systematically obtained from the Hamiltonian by means of its symbol $\mathcal{H} (\boldsymbol{k}, \boldsymbol{r}) $, an operator  which generalises the Bloch Hamiltonian and contains all topological information. This approach, when applied to Dirac operators, allows to recover the tenfold classification of insulators and superconductors. The existence of a stable $\mathbb{Z}$-topology is predicted as a condition on the dimension $D$, similar to the classification of defects in thermodynamic phase transitions. Kekule distortions, vacancies and adatoms in graphene are proposed as examples of such defects and their topological equivalence is discussed.

\end{abstract}

\pacs{Valid PACS appear here}% PACS, the Physics and Astronomy
\maketitle

\section{Introduction}
The tenfold classification of insulators and superconductors 
generalises the Dyson classification for disordered systems \cite{Altland1997} and proposes a systematic scheme to identify topological features. It is based on anti-unitary symmetries, time reversal (TRS) and particle-hole (PHS), the unitary chiral symmetry (CS) and the space dimension $d$ of non-interacting systems of fermions  \citep{Kitaev2009,Schnyder2008}.  
%based on Clifford algebras of Dirac matrices \citep{Kitaev2009,Schnyder2008}
%The search for topological classes involves powerful methods and elegant topics in mathematics and in condensed matter physics.
%, such as IQHE, Berry and Zak phases, and topologically invariant characteristics. 
%Yet, despite spectacular progresses, the tenfold classification remains challenging, partly due to the difficulty to decipher some of the underlying mathematical concepts and make them accessible to directly observable  quantities. 
 %This generalisation is based on the role of the antiunitary particle-hole  symmetry (PHS) in addition to the time reversal (antiunitary) symmetry (TRS). Their product, the chiral symmetry (CS), is unitary but anticommutes with the Hamiltonian. 
%Equipped with the 10 possible combinations of these three symmetries, a tenfold classification %applicable to finite dimensions $d$ 
%has been proposed, based on Clifford algebras of Dirac matrices \citep{Kitaev2009,Schnyder2008}. 
For each symmetry class, topological properties are deduced from band structures of Bloch Hamiltonians using a homotopy group based on the powerful yet abstruse $K$-theory classifying fibre bundles hence topological structures \citep{Kitaev2009,Stone2010}. The principle leading to this classification remains challenging, and  
it hinders %to some extent 
the role played by defects and disorder. This latter aspect has been considered \cite{Teo2010} and the tenfold classification has been modified replacing $d$ by 
\begin{equation}
\label{eq:deltaDefinition}
\delta \equiv d - D,
\end{equation}
where the dimension $D$ characterizes the spatial envelope of defects.

This paper offers an alternative approach in the spirit of defect classification in thermodynamic phase transitions \cite{Toulouse1976}. Introducing the tunable dimension $D$, we show how to change topological features of a given symmetry class in the tenfold classification by creating spatial defects or textures. We mostly consider examples from the $BDI$ symmetry class, essentially monolayer graphene, an always surprising system, whose Bloch Hamiltonian,  
gapless at Dirac points, displays the three symmetries (TRS, PHS and CS). In two dimensions, this class is not topological (see Table \ref{tab:10-fold_way}). 
Creating defects, e.g. atomic vacancies, adatoms or modulated perturbations, we discuss the conditions on $D$, so as to turn monolayer graphene into a topological material with non vanishing integer invariant (a Chern or a winding number \citep{KOHMOTO1985, TKNN1982}) and topologically protected (zero modes) edge states. 
%We present an explicit calculation of their $\pm 1$ winding numbers and a way to directly access them  experimentally. 
We show that defects and textures can be grouped into universality classes characterised by 4 parameters: two dimensions $(d,D)$ and two integers $(p,q)$ counting the number of Dirac matrices assigned to a Hamiltonian with specific defects. With (\ref{eq:conditions}), we prove a necessary condition to observe  topological features .
This condition applies to the 8 real symmetry classes of Table \ref{tab:10-fold_way}, and it allows to retrieve all $\mathbb{Z}$-integer classes. When applied to the $BDI$ symmetry class, (\ref{eq:conditions})  indicates that graphene  with a vacancy belongs to the same universality class as a Kekule distorsion, a rich model which displays fractional charge, $\mathbb{Z}$-topology and a clear illustration of the Atiyah-Singer Index theorem \citep{Atiyah1963, Atiyah1968, Nakahara1990, Stone1984, Chiu2016, Callias1978, Ertem2017, Fukui2012, Eguchi1980, Niemi&Semenoff1984, Getzler1983, Akkermans1998}. Other results are summarised in Table \ref{tab:topological_systems}. 
The paper is organized as follows. In section \ref{sec:symbol} we define the symbol of a Hamiltonian and discuss its topological content. In section \ref{sec:10-fold way}, we use the symbol to constructively establish the 10-fold classification. In section \ref{sec:defects}, we present examples of topological defects. In section \ref{sec:index thorem} we relate topological invariant numbers to edge states using an index theorem. Section \ref{sec:no_AS} discusses the applicability conditions of the index theorem, while section \ref{sec:conclusion} summarizes our results.
\section{The symbol operator}
\label{sec:symbol}
The simplest situation where both TRS and PHS can be implemented is a two-band model. Translation symmetry and Bloch theorem allow the reduction of a tight binding Hamiltonian matrix $H$ to a $2 \times 2$ Bloch Hamiltonian matrix $\mathcal{H}(\boldsymbol{k}) $. The pseudo-momentum $\boldsymbol{k}$ takes values in the Brillouin zone, a $d$-dimensional compact torus $\mathbb{T}^d$. The energy spectrum of $H$ is retrieved from the band energy spectrum $E_n ( \boldsymbol{k} )$ of $\mathcal{H}(\boldsymbol{k})$ and the topological properties obtained from the Berry connection  are encoded in the Bloch wavefunctions.
The tenfold classification relies upon a generalisation of this model using Clifford algebras of $n \times n$ anticommuting Dirac matrices. 
%Maps between the Brillouin zone $\mathbb{T}^d$ to a target Hilbert space $\mathbb{C}^n$ of the Bloch Hamiltonian allow to identify and calculate topologically invariant quantities. For two bands $(n=2, m=1)$, the projective space $\mathbb{C}P^1$ can be viewed as $U(1)$-invariant wavefunctions defined on the two-dimensional sphere $S^2$. Topological properties are encoded in the map from $\mathbb{T}^d$ to $S^2$ hence depending on the spatial dimension $d$ and on $m$. 
%This approach makes use of translation symmetry absent in $d=0$ random systems initially considered \cite{Altland1997}.
%, and more generally, when a spatially varying potential prevents using the Bloch theorem. 
In the presence of a spatially varying and non periodic potential  \citep{Teo2010}, we associate to a Hamiltonian $H(-i \, {\boldsymbol \nabla} , \boldsymbol{r})$ its symbol $\mathcal{H} (\boldsymbol{k}, \boldsymbol{r}) $  defined by the Weyl transform \cite{Hillery1984, Case2008},
\begin{equation}
\label{eq:Weyl_transform}
\small
\mathcal{H} (\boldsymbol{k}, \boldsymbol{r})  = \int_{-\infty} ^\infty d\boldsymbol{r}^\prime e^{-i\boldsymbol{k}\cdot \boldsymbol{r}^\prime} \bra{\boldsymbol{r}+\frac{\boldsymbol{r}^\prime}{2}}H\ket{\boldsymbol{r}-\frac{\boldsymbol{r}^\prime}{2}} \, .
\end{equation}
The symbol is routinely used when translation symmetry allows replacing $- i \partial_j $ by a wavenumber $k_j$, hence recovering the Bloch Hamiltonian, namely
\begin{equation}
\small
\label{eq:symbol_bloch}
\begin{aligned}
\mathcal{H}\left(\boldsymbol{k}\right) & =\int_{-\infty}^{\infty}d\boldsymbol{r}^{\prime}\,e^{-i\boldsymbol{k}\boldsymbol{r}^{\prime}}\left\langle \boldsymbol{r}+\frac{\boldsymbol{r}^{\prime}}{2}|H\left(\hat{k}\right)|\boldsymbol{r}-\frac{\boldsymbol{r}^{\prime}}{2}\right\rangle \\
 & =\frac{1}{2\pi}\int_{-\infty}^{\infty}d\boldsymbol{r}^{\prime}d\boldsymbol{k}^{\prime}\,e^{-i\boldsymbol{k}\boldsymbol{r}^{\prime}}e^{i\boldsymbol{k}^{\prime}\boldsymbol{r}^{\prime}}H\left(\boldsymbol{k}^{\prime}\right)\\
 & =\int_{-\infty}^{\infty}d\boldsymbol{k}^{\prime}H\left(\boldsymbol{k}^{\prime}\right)\delta\left(\boldsymbol{k}-\boldsymbol{k}^{\prime}\right)\\
 & =H\left(\boldsymbol{k}\right).
\end{aligned}
\end{equation}
 The correspondence between a Hamiltonian and its symbol has been formalised \cite[Chapter ~24-27]{yankowsky2013} and used to quantize classical systems by means of a Wigner transform,
\begin{equation}
\label{eq:wignerTransform}
\small
H(\Hat{\boldsymbol{k}},\Hat{\boldsymbol{r})} = \frac{1}{(2\pi)^2}\int_{-\infty} ^\infty d\boldsymbol{a} \, d\boldsymbol{b} \, \Tilde{\mathcal{H}}(\boldsymbol{a},\boldsymbol{b}) \, e^{i\left(\boldsymbol{a}\cdot \Hat{\boldsymbol{r}}+\boldsymbol{b}\cdot \Hat{\boldsymbol{k}}\right)},
\end{equation}
where $\Tilde{\mathcal{H}}(\boldsymbol{a},\boldsymbol{b})$ is the Fourier transform of $\mathcal{H} (\boldsymbol{k}, \boldsymbol{r})$ with respect to $\boldsymbol{k}$ and $\boldsymbol{r}$.
%These transformations are routinely used, the Wigner transform of a classical Hamiltonian is the Weyl ordered quantum Hamiltonian, while the Weyl transform of the density matrix gives the Wigner function.
A Hamiltonian and its symbol do not generally share the same spectrum, e.g., a harmonic oscillator and its symbol have respectively a discrete (quantum) and  a continuous (classical) spectra. Yet, in the presence of translation symmetry, a unitary transformation relates a  
Hamiltonian to its symbol, the Bloch Hamiltonian $\mathcal{H}(\boldsymbol{k})$ so that both can be used  interchangeably. 
%    H = \sum_{\boldsymbol{k}}a^{\dagger}_{\boldsymbol{k}}\mathcal{H}(\boldsymbol{k})a_{\boldsymb%ol{k}}
%\end{equation}
%The Bloch Hamiltonian $\mathcal{H}(\boldsymbol{k})$ is
%a $n \times n$ matrix, where $n$ is the number of atoms in the unit cell. 
%The pseudo-momentum $\boldsymbol{k}$ becomes a good quantum number, namely  
%translation symmetry allows to use the .
This state of affairs is the exception rather than the rule and although Bloch theorem does not apply when translation symmetry is broken, the symbol $\mathcal{H} (\boldsymbol{k}, \boldsymbol{r})$ is well defined. It does no longer provide information on the Hamiltonian spectrum, but it encodes all topological information \citep{Atiyah1963,Atiyah1968, Nakahara1990}, \cite[Chapters~26-27]{yankowsky2013}. 
 Topologically invariant numbers can be retrieved from eigenfunctions of $\mathcal{H} (\boldsymbol{k}, \boldsymbol{r}) $ provided a spectral gap exists in its  eigenvalue spectrum. Note that the existence of this gap does not imply a similar behaviour of the Hamiltonian except in the presence of translation symmetry. This remark is of utmost importance and it will become much  clearer in the following example.
 
 Graphene \cite{Neto2009}, a bipartite honeycomb lattice of carbon atoms well described by the Hamiltonian 
\begin{equation}
\label{eq:graphene_NN_tb}
    H_{0} = -t\sum_{i}\sum_{\delta=0}^2 a^{\dagger}_ib_{i+\delta}^{\vphantom{\ast}} +h.c. \, ,
\end{equation}
with hopping $t$ between nearest neighbors, provides a relevant  playground to test these ideas. $H_{0}$ is diagonal in momentum space 
% \footnote{
% \begin{equation}
% \label{eq:graphene_momentum_tb}
% H_{0} = \sum_{\boldsymbol{k}}\left(\begin{array}{cc}
% a^\dagger_{\boldsymbol{k}} &
% b^{\dagger}_{\boldsymbol{k}}
% \end{array}\right)\mathcal{H}(\boldsymbol{k})\left(\begin{array}{c}
% a_{\boldsymbol{k}}\\
% b_{\boldsymbol{k}}
% \end{array}\right)
% \end{equation}}  
and its energy spectrum displays two independent Dirac points, $K$ and $ K^\prime$.
%, a general result known as fermion doubling \citep{NIELSEN1981a,NIELSEN1981b}.
At low energy and in appropriate units, the Bloch Hamiltonian associated to $H_0$ is
\begin{equation}
\label{eq:graphene_bloch_dirac}
\mathcal{H}_0(\boldsymbol{k}) = k_x \, \sigma_x\otimes\tau_z + k_y \, \sigma_y \otimes \mathbf{1} 
\end{equation}
in the sublattice basis $ \psi_{\boldsymbol{k}} = \left(\begin{array}{cccc}
a^K_{\boldsymbol{k}} &
a^{K^\prime}_{\boldsymbol{k}} &
b^K_{\boldsymbol{k}} &
b^{K^\prime}_{\boldsymbol{k}}
\end{array}\right)^t$ with $\boldsymbol{k} = (k_x , k_y)$. %Graphene is a physical realization of a Dirac Hamiltonian with a gapless spectrum at Dirac points. 
In %the $4 \times 4$ matrices 
$\sigma_i\otimes\tau_j$, the Pauli matrix $\sigma_i$ (resp. $\tau_j$) corresponds to the sublattice $(A,B)$ (resp. the valley $K,K'$) degrees of freedom.
The symbol (\ref{eq:graphene_bloch_dirac}) is invariant under antiunitary symmetries (TRS) and (PHS) and chiral symmetry, $\{\mathcal{H}_0(\boldsymbol{k}) , \sigma_z \otimes \mathbf{1}\} =0$. Hence it belongs to the $BDI$ class which, for $d =2$, does not display topological features (see Table \ref{tab:10-fold_way}).
%The simple implication of these two Dirac points is that graphene is gapless and therefore cannot be classified in the 10-fold way as it requires a gap. If we ignore this problem and try to still classify it in the 10-fold way we will find that it is topologically trivial as we shall see later on. 
 %The expression (\ref{2by2}) of the Hamiltonian does not account for all type of spatial perturbation. A simple example for this situation is provided by the case of an atomic vacancy, namely the removal of a single atom in the graphene honeycomb lattice. 
%We consider a non interacting electron gas on a lattice in $\mathbb{R}^d$ that we describe by means of a tight binding Hamiltonian ${\cal H} = \sum_{\langle i,j \rangle} t_{ij} |i \rangle \langle j |$, with nearest neighbour hopping energies $t_{ij}$. For a translation invariant lattice, the Bloch theorem allows to write 
%Usually topological features are invoked when looking at the dynamics in a single band and the Berry curvature appears to describe virtual transitions between bands. Hence topology shows up when calculating physical quantities related to single band dynamics. This is the case for the Hall conductance according to TKNN. 

\begin{figure}[ht]
    \centering
    \includegraphics[width=0.38\textwidth]{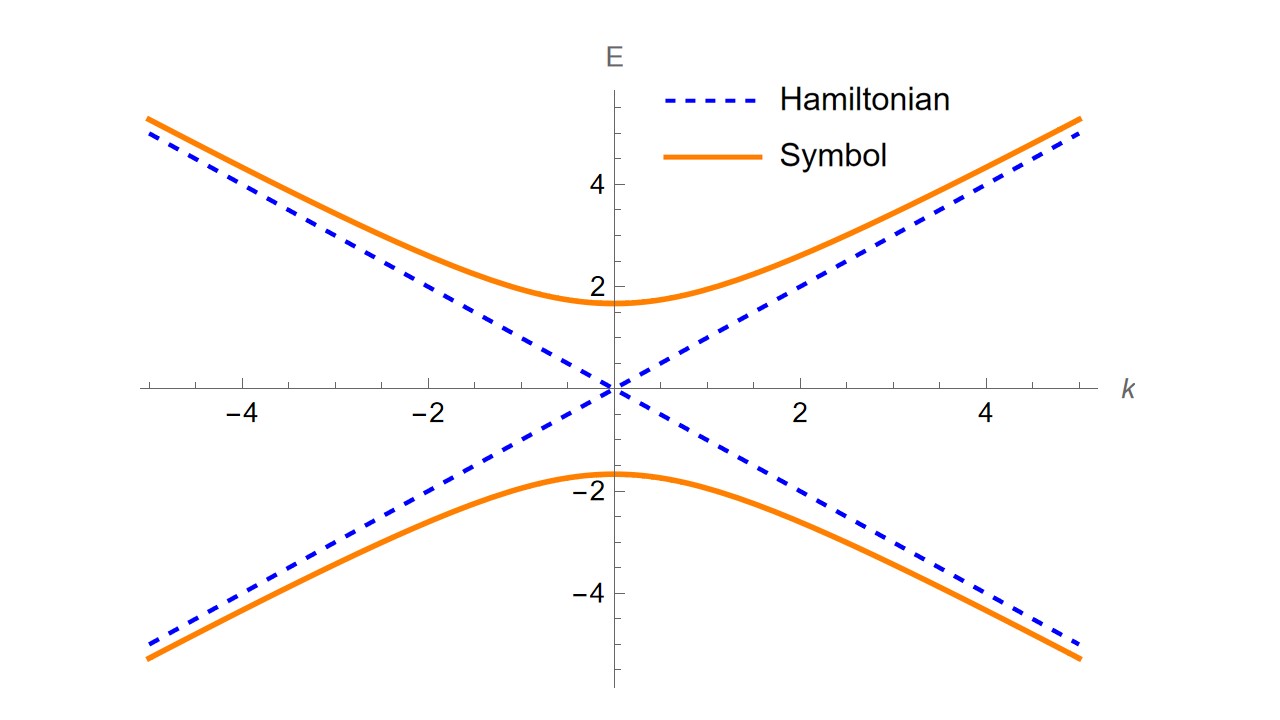}
	\caption{Spectral behaviour of graphene with a vacancy. The dashed line corresponds to the spectrum of the Hamiltonian $H_V$ in (\ref{eq:graphene_NN_tb_vacancy}). It is linear with $\boldsymbol{k}$ and gapless. The full line corresponds to the spectrum of the symbol  (\ref{eq:H_graphene}) with $\phi (r) = \frac{1}{r\ln{{r}/{r_0}}}$.}
	\label{fig:Hamiltonian_Symbol_spectrum}
\end{figure}

Consider now creating a vacancy \citep{Kelly1998, Ugeda2010, Ovdat2017, Pereira2006, Pereira2008, Amara2007, Palacios2008, Dutreix2013} by the removal of a neutral carbon atom located at the origin of the lattice. Hamiltonian (\ref{eq:graphene_NN_tb})  becomes, 
\begin{equation}
\label{eq:graphene_NN_tb_vacancy}
H_{V} = -t\sum_{i}\sum_{\delta=0}^2 a^{\dagger}_ib_{i+\delta}^{\vphantom{\ast}} + t\sum_{\delta=0}^2 a^{\dagger}_0b_{0+\delta}^{\vphantom{\ast}} +h.c.
\end{equation}
The vacancy breaks translation symmetry so that $\boldsymbol{k}$ is no longer a good quantum number, but $H_V$ can still be expanded at low energy around the Dirac points,
%. In spatial representation,
\begin{equation}
\label{eq:graphene_d_vacancy}
H_V = \int \, d\boldsymbol{r} \, 
\psi^{\dagger}_{\boldsymbol{r}} \, 
{H}(\boldsymbol{r},-i \, \partial\boldsymbol{_r}) \, 
\psi_{\boldsymbol{r}} \, .
\end{equation}
%where $\psi_{\boldsymbol{r}} = \left(\begin{array}{cccc}
%a^K_{\boldsymbol{r}} &
%a^{K^\prime}_{\boldsymbol{r}} &
%b^K_{\boldsymbol{r}} &
%b^{K^\prime}_{\boldsymbol{r}}
%\end{array}\right)^t$ and
Making use of the Weyl transform (\ref{eq:Weyl_transform}), we obtain the symbol,
%, the symbol of $H_V$, is 
\begin{equation} \label{eq:H_graphene} \begin{aligned}
\mathcal{H}_V (\boldsymbol{k}, \boldsymbol{r}) & =k_{x}\, \sigma_{x}\otimes\tau_{z}+k_{y} \, \sigma_{y}\otimes \mathbf{1} \\ & + {\phi}_{1} \left(\boldsymbol{r} \right)\sigma_{x}\otimes\tau_{x}+ \phi_{2}\left(\boldsymbol{r} \right)\sigma_{x}\otimes\tau_{y}\\
%  & =\left(\begin{array}{cccc}
% 0 & 0 & k_{x}-ik_{y} & \phi e^{-i\Delta}\\
% 0 & 0 & \phi e^{i\Delta} & k_{x}+ik_{y}\\
% k_{x}+ik_{y} & \phi e^{-i\Delta} & 0 & 0\\
% \phi e^{i\Delta} & k_{x}-ik_{y} & 0 & 0
% \end{array}\right)
\end{aligned} 
\end{equation}
with 
$\boldsymbol{\phi} (\boldsymbol{r})  \equiv \phi_{1}+i \phi_{2}= \phi (r) \, e^{-i\theta}$, and  $\theta$ being the angle relative to $\boldsymbol{r}$ (see appendix \ref{sec:app_a}).
%This illustrates the previous remark about the gap. 
A vacancy does not open a gap in the $k$-linear spectrum of $H_V$ but it creates zero modes at the Fermi energy %taken to be zero 
(for undoped graphene) and it couples the valleys \citep{Pereira2008, Mallet2012, Ovdat2020}. However, the eigenvalue spectrum of the symbol (\ref{eq:H_graphene}) has a gap for any non vanishing function $\phi(r) = (\phi_1^2+ \phi_2^2)^{1/2}$.
% \footnote{Consider for instance the function $\phi(r) = \frac{1}{r\ln{\frac{r}{r_0}}}$, where $r_0$ is an arbitrary constant, e.g. the lattice constant. This is a non-vanishing function of $r$ so that the symbol has a gap.}
%but the Hamiltonian has a  so it is gapless, see the appendix \ref{app:H^2_graphene_vacancy} for a derivation of this result.
This difference between a Hamiltonian and its symbol, is displayed in Fig \ref{fig:Hamiltonian_Symbol_spectrum}. Graphene with a vacancy does not have a spectral gap hence it apparently does not belong to the tenfold classification. But its symbol has a gap and it is what matters for topological properties and the tenfold classification. This point is usually overlooked since in the presence of translation symmetry, a Hamiltonian and its symbol have related spectra.

Both $H_V$ and $\mathcal{H}_V (\boldsymbol{k}, \boldsymbol{r}) $ are invariant under the antiunitary symmetries (TRS), (PHS) and chiral symmetry, $\{\mathcal{H}_V (\boldsymbol{k}, \boldsymbol{r}) , \sigma_z\otimes \mathbf{1} \} =0$, so that graphene with a vacancy also belongs to $BDI$, seemingly  void of topological features for $d =2$. We wish to revisit that result. 

\section{Building the 10-fold way with the symbol}
\label{sec:10-fold way}
We present an alternative derivation of the tenfold classification of insulators and superconductors based  on properties of symbols $\mathcal{H}\left(\boldsymbol{k},\boldsymbol{r}\right)$.   
We restrict our discussion to the 8 real symmetry classes displaying at least one of the antiunitary symmetries (TRS) or (PHS). We recall for convenience that they are respectively described by $T = U_{T}K$ and $P = U_{P}K$ where $U_{T},U_{P}$ are unitary operators and $K$ is the anti-unitary complex conjugate. The chiral symmetry $C = TP=U_{T}U_{P}^{*}$ is their product.
These symmetries translate for the symbols $\mathcal{H}\left(\boldsymbol{k},\boldsymbol{r}\right)$ into the requirements,
\begin{equation} \label{eq:symmetriesHamiltonian} \begin{aligned}
T\mathcal{H}\left(\boldsymbol{k},\boldsymbol{r}\right)T^{-1} & =\mathcal{H}\left(-\boldsymbol{k},\boldsymbol{r}\right)\\
P\mathcal{H}\left(\boldsymbol{k},\boldsymbol{r}\right)P^{-1} & =-\mathcal{H}\left(-\boldsymbol{k},\boldsymbol{r}\right)\\
C\mathcal{H}\left(\boldsymbol{k},\boldsymbol{r}\right)C^{-1} & =-\mathcal{H}\left(\boldsymbol{k},\boldsymbol{r}\right) \, .
\end{aligned} 
\end{equation}
%\begin{table}[ht]
%\centering
%\begin{adjustbox}
%{width=0.40\textwidth}
%\begin{center}
\newlength\celldim
\newlength\fontheight
\newlength\extraheight
\setlength\celldim{2.25em}
\settoheight\fontheight{A}
\setlength\extraheight{\celldim - \fontheight}
\newcolumntype{S}
{ @{}
>{\centering\arraybackslash}
p{\celldim}
<{\rule[-0.2\extraheight]{0pt}%
{\fontheight + \extraheight/4}}
@{} }

\begin{table}
\small
\begin{center}\begin{tabular}{|S|S|S|S|S|S|S|S|S|}
\hline 
\text{Class} & $s$ & $T$ & $P$ & $C$ & $\delta=0$ & 1 & 2 & 3
\\
\hline 
\hline 
A & 0 & 0 & 0 & 0 & \cellcolor{cyan}$\mathbb{Z}$ & 0 & \cellcolor{cyan}$\mathbb{Z}$ & 0
\\
\hline 
AIII & 1 & 0 & 0 & 1 & 0 & \cellcolor{cyan}$\mathbb{Z}$ & 0 & \cellcolor{cyan}$\mathbb{Z}$
\\
\hline 
\hline
AI & 0 & + & 0 & 0 & \cellcolor{cyan}$\mathbb{Z}$ & 0 & 0 & 0
\\
\hline 
BDI & 1 & + & + & 1 & \cellcolor{orange}$\mathbb{Z}_2$ & \cellcolor{cyan}$\mathbb{Z}$ & 0 & 0
\\
\hline 
D & 2 & 0 & + & 0 & \cellcolor{orange}$\mathbb{Z}_2$ & \cellcolor{orange}$\mathbb{Z}_2$ & \cellcolor{cyan}$\mathbb{Z}$ & 0
\\
\hline 
DIII & 3 & - & + & 1 & 0 & \cellcolor{orange}$\mathbb{Z}_2$ & \cellcolor{orange}$\mathbb{Z}_2$ & \cellcolor{cyan}$\mathbb{Z}$
\\
\hline 
AII & 4 & - & 0 & 0 & \cellcolor{green}$2\mathbb{Z}$ & 0 & \cellcolor{orange}$\mathbb{Z}_2$ & \cellcolor{orange}$\mathbb{Z}_2$
\\
\hline 
CII & 5 & - & - & 1 & 0 & \cellcolor{green}$2\mathbb{Z}$ & 0 & \cellcolor{orange}$\mathbb{Z}_2$
\\
\hline 
C & 6 & 0 & - & 0 & 0 & 0 & \cellcolor{green}$2\mathbb{Z}$ & 0
\\
\hline 
CI & 7 & + & - & 1 & 0 & 0 & 0 & \cellcolor{green}$2\mathbb{Z}$
\\
\hline
\end{tabular}
\end{center}
\caption{Tenfold classification. The first five columns display the 10 symmetry classes labeled by $s$ and  defined by their antiunitary symmetries $T,P$ and chirality $C$. "$+$" ("$-$") means that the relevant operator is a symmetry which squares to $1$ $(-1)$ and $"0"$ the absence thereof. The last 4 columns indicate possible topological classes $(0, \mathbb{Z}, \mathbb{Z}_2 ) $ as a function of the reduced dimension $\delta = d - D$. \label{tab:10-fold_way}}
\end{table}
Table \ref{tab:10-fold_way} accounts for  symmetry classes of symbols $\mathcal{H}\left(\boldsymbol{k},\boldsymbol{r}\right)$ of generic form, 
\begin{equation}
%\small
\label{eq:Diracdefnition}
\mathcal{H}\left(\boldsymbol{k},\boldsymbol{r}\right)=\boldsymbol{h}_s \cdot\boldsymbol{\gamma}_s+ \, \boldsymbol{h}_a \cdot\boldsymbol{\gamma}_a \equiv \, \boldsymbol{h}\left(\boldsymbol{k},\boldsymbol{r}\right) \cdot \boldsymbol{\gamma} \, .
\end{equation}
The table extends to $\delta = 7$ and it is periodic both in $s$ and $\delta$. This constitutes Bott periodicity. For a discussion of this periodicity and its  dependence on $\delta$, we direct the reader to \cite{Teo2010}.
The set of anticommuting $\boldsymbol{\gamma}$ matrices is conveniently split into position and momentum like, $\boldsymbol{\gamma}_s$ and $\boldsymbol{\gamma}_a$ matrices according to (\ref{eq:symmetriesHamiltonian}), viz. symmetric $\boldsymbol{h}_s$ or antisymmetric $\boldsymbol{h}_a $ fields, 
\begin{equation} \label{eq:DiracSymmetries}
\begin{aligned}
\boldsymbol{h}_s\left(\boldsymbol{k},\boldsymbol{r}\right) & =\boldsymbol{h}_s\left(-\boldsymbol{k},\boldsymbol{r}\right),&&T\gamma_s T^{-1}=\gamma_s \\
\boldsymbol{h}_a\left(\boldsymbol{k},\boldsymbol{r}\right) & =-\boldsymbol{h}_a\left(-\boldsymbol{k},\boldsymbol{r}\right),&&T\gamma_a T^{-1}=-\gamma_a \, .
\end{aligned} 
\end{equation}
This implies 
\begin{equation}
\label{eq:symmetriesCommutations}
\begin{aligned}
\left[\gamma_s,T\right] & =\left\{ \gamma_a,T\right\} =0 \\
\left\{ \gamma_s,P\right\} & =\left[\gamma_a,P\right]=0\\
\left\{ \gamma_s,C\right\} & =\left\{ \gamma_a,C\right\} =0.
\end{aligned}
\end{equation}
We denote $p$ (resp. $q+1$) the number of $\gamma_a$ (resp. $\gamma_s$) matrices. The momentum $\boldsymbol{k}$ depends on the spatial dimension $d$ of the lattice and the position  
$\boldsymbol{r}$ which accounts for a potential breaking  translation symmetry  (hereafter a defect), is  characterised by a dimension $D$. Hence symbols $\mathcal{H}\left(\boldsymbol{k},\boldsymbol{r}\right)$ in (\ref{eq:Diracdefnition}) depend on the parameters $(d,D,p,q)$ and are classified by $s=p-q\;(\text{mod}\,8)$. A scaling relation, 
\begin{equation}
\label{eq:s-delta} 
\mathcal{H} (d,D,p,q) = \mathcal{H}(s - \delta) \, ,    
\end{equation} 
holds between $s$ and $\delta \equiv d - D $  
\cite{Teo2010} (see appendix \ref{sec:app_b}).

%---- Toplogical numbers ------%

$\mathbb{Z}$-topology classes in Table \ref{tab:10-fold_way} correspond to integer values of invariant integrals encoded in the eigenfunctions of the Bloch Hamiltonian. 
%defining a Berry connection $\boldsymbol A$. If it is a closed form, $d \boldsymbol A =0$, we check if it is exact, ${\boldsymbol A} = d \boldsymbol \omega$. If so, its integral over a closed trajectory in phase space vanishes due to Stokes theorem and the phase space is topologically trivial. However, if $\boldsymbol A$ is not exact, this integral is a nonzero integer expressing topological obstruction. 
This result also holds for symbols $\mathcal{H}\left(\boldsymbol{k},\boldsymbol{r}\right)$.  
The $(p + q +1) $-component field $\boldsymbol{h} (\boldsymbol{k} , \boldsymbol{r})$ in (\ref{eq:Diracdefnition}) can be normalised, ${\boldsymbol{\hat h}} = \boldsymbol{h} / h$  since the spectrum of the symbol has a gap ($h \neq 0$), hence we define,
\begin{equation}
\label{eq:linearDiracGeneral}
 {\mathcal{\hat H}}\left(\boldsymbol{k},\boldsymbol{r}\right)=  {\boldsymbol{\hat h}} \left(\boldsymbol{k},\boldsymbol{r}\right) \cdot \boldsymbol{\gamma} \,  , 
\end{equation}
a procedure often called flattening. Each component   
$\hat{h}^i$ depends on $d+D$ variables, the $d$ components $(k_1, \dots, k_d)$ of $\boldsymbol{k}$ on the $S^d$ sphere
% \footnote{We recall that the absence of translation symmetry prevents defining a quasi-momentum in the Brillouin zone torus $\mathbb{T}^d$. }
and the $D$ spatial components $\boldsymbol{r} = (r_1, \dots, r_D)$ of the defect, defined on $S^D$. For Dirac symbols (\ref{eq:Diracdefnition}), $\boldsymbol{\gamma}$ are $n \times n$ Dirac matrices with $n = 2^m$.
\subsection{Topological Invariants}
There exist two types of topologically invariant integers, depending on the parity of $d+D$. For even (odd) values, a Chern number $\mathcal{C}$ (a winding number $\nu$) is defined to capture phase space topological obstructions.
They are %For even (odd) values of $m$, the corresponding Chern (winding) numbers are 
expressed by the integrals,
\begin{equation} \label{eq:chernDefinition} 
\mathcal{C}_{m} \propto\int_{S^{d+D}} \, 
%d^d k \, d^D r\, 
\text{Tr}\left[\hat{\mathcal{H}}\left(d\hat{\mathcal{H}}\right)^{2m}\right]
\end{equation}
where $(\text{Tr})$ is over the $\boldsymbol{\gamma}$ matrices and for chiral symbols,  
$\hat{\mathcal{H}} = {\boldsymbol{\hat h}} \cdot \boldsymbol{\gamma} =\begin{psmallmatrix}
        0 & \cal{Q} \\
        \cal{Q}^{\dagger} & 0
    \end{psmallmatrix} $,
    % \footnote{Since  $d\left(\hat{\mathcal{H}}\left(d\hat{\mathcal{H}}\right)^{2m}\right) = d \left(q^\dagger dq\right)^{2m-1} = 0$. The prefactor in these integrals is not important at this stage. We will later simplify these expressions.}
\begin{equation} \label{eq:windingDefinition} 
\nu_{2m-1} \propto\int_{S^{d+D}} \, 
%d^d k \, d^D r\,
\text{Tr}\left[\left({\cal{Q}}^{\dagger} d \cal{Q}\right)^{2m-1}\right] \, .
\end{equation} 
For non chiral symbols, using the Clifford decomposition (\ref{eq:linearDiracGeneral}), 
${\mathcal{\hat H}}\left(\boldsymbol{k},\boldsymbol{r}\right)=  {\boldsymbol{\hat h}} \cdot \boldsymbol{\gamma} $, 
we have, 
\begin{equation} 
\small
\begin{aligned}
\label{eq:chernDirac} 
 \mathcal{C}_{m} & \propto\int\text{Tr}\left[\mathcal{\hat H}\left(d\mathcal{\hat H}\right)^{2m}\right]\\ & =\int\sum_{i_{1},i_{2},\cdots i_{2m+1}} h_{i_{1}}dh_{i_{2}}\cdots dh_{i_{2m+1}} \text{Tr}\left[\gamma_{i_{1}}\gamma_{i_{2}}\cdots\gamma_{i_{2m+1}}\right] \, .
\end{aligned}
\end{equation}
This expression does not vanish for $i_{2}\neq i_{3}\neq\cdots\neq i_{2n+1}$, hence  $\gamma_{i_{2}}\neq\gamma_{i_{3}}\neq\cdots\neq\gamma_{i_{2n+1}}$.
A Dirac symbol without chiral symmetry must be built out of all possible anti-commuting $\gamma$ matrices whose product is 
proportional to the identity. This implies that $\mathcal{C}_{m}$ in expression (\ref{eq:chernDirac}), 
does not vanish for  $\gamma_{i_{1}}\neq\gamma_{i_{2}}\neq\gamma_{i_{3}}\neq\cdots\neq\gamma_{i_{2m+1}}$.
Since  $\gamma$ matrices anti-commute, exchanging indices will
give relative $\pm$ signs depending on the exchanges, overall
this expression gives complete anti-symmetrization of all indices
and is nothing but an integral of the Jacobian,
\begin{equation} 
\label{eq:DiracDeterminant}
\small
J (\boldsymbol{h}, d, D) = \left|\begin{array}{cccc}
h_{1} & h_{2} & \cdots & h_{d+D+1}\\
\partial_{1}h_{1} & \partial_{1}h_{2} & \cdots & \partial_{1}h_{d+D+1}\\
\vdots & \vdots & \ddots & \vdots\\
\partial_{d+D}h_{1} & \partial_{d+D}h_{2} & \cdots & \partial_{d+D}h_{d+D+1}
\end{array}\right|,
\end{equation}
provided we use the relation $2m = d+D$.

For chiral classes, we calculate a winding number written under the form,
\begin{equation} \label{eq:DiracWinding} \begin{aligned}
\nu_{2m-1} & \propto\int\text{Tr}\left[\left(\mathcal{Q}^{\dagger}d\mathcal{Q}\right)^{2m-1}\right]\\
 & =\left(-1\right)^{m-1}\int\text{Tr}\left[\mathcal{Q}^{\dagger}d\mathcal{Q}\left(d\mathcal{Q}^{\dagger}d\mathcal{Q}\right)^{m-1}\right]
\end{aligned} 
\end{equation}
where we have used the unitarity of $\mathcal{Q}$. In a basis where the symbol $\mathcal{\hat H}=\left(\begin{array}{cc}
0 & \mathcal{Q}\\
\mathcal{Q}^{\dagger} & 0
\end{array}\right)=\hat{\boldsymbol{h}}\cdot\boldsymbol{\gamma}$, one $\gamma$ matrix is necessarily given either  by $\gamma_{1}=\sigma_{x}\otimes\unit$
or $\gamma_{1}=\sigma_{y}\otimes\unit$ and the rest are  given by
$\gamma_{i}=\sigma_{y}\otimes\tilde{\gamma}_{i}$ or $\gamma_{i}=\sigma_{x}\otimes\tilde{\gamma}_{i}$ accordingly, 
 where $\tilde{\gamma}_{i}$'s now define a complete set of
anti-commuting $\gamma$ matrices. For simplicity,  assume that $\gamma_{1}=\sigma_{y}\otimes\unit$ hence, 
\begin{equation} \label{eq:general_q_flattened} 
\mathcal{Q}=-ih_{1}\unit{+}\sum_{i=2}^{2m-1}h_{i}\tilde{\gamma}_{i}\equiv \hat{\boldsymbol{h}}\cdot\tilde{\boldsymbol{\gamma}} \, ,
\end{equation}
a situation identical to the previous case of a Chern number. The product of all $\tilde{\gamma}$'s still being  proportional to
the identity matrix, the winding number is again proportional
to an integral of the determinant (\ref{eq:DiracDeterminant}), provided we use the relation $2m-1 = d+D$.
This determinant does not vanish only if the number
of variables in $(\boldsymbol{k}, \boldsymbol{r})$ and in $\boldsymbol{h}$ are identical, namely if
\begin{equation}
\label{eq:conditions}
\small
d + D = p + q \, .
\end{equation} 
Overall, fixing a prefactor $S_{d+D}$, the surface of the unit sphere in $(d+D)$ dimensions, both Chern and winding numbers are given by an integral over the same determinant (\ref{eq:DiracDeterminant}),
% \footnote{As an immediate consequence, non chiral classes have integer topological numbers in even dimensions and chiral classes in odd dimensions. A
% chiral class has $s\quad\text{mod}\:2=1$ so $p-q$ must be odd and
% therefore $p+q=p-q+2q$ is also odd. Therefore, to have non vanishing
% integer topological number $d+D$ must also be odd. So chiral classes
% only have winding numbers as integer topological numbers while non
% chiral classes have Chern numbers.}
\begin{equation}
\label{eq:chern_winding_determinant}
\nu_{d+D} = \frac{1}{S_{d+D}}\int_{S^{d+D}}d^dk \, d^Dr\, J(\boldsymbol{h}, d, D) \, ,
\end{equation}
for all classes $s$ and dimensions $(d+D)$.  
The only difference is that $d + D = 2m -1$ for chiral classes, i.e. for odd values of $s = p-q$, while $d + D = 2m$ for non chiral classes i.e. for even values of $s$. 
This result can still be used for non normalised (flattened) symbols, provided we introduce a prefactor  $(f^{d+D+1})^{-1}$ in the Jacobian, where 
\begin{equation}
\label{eq:prefactor}
f=\sqrt{h_{1}^{2}+h_{2}^{2}+\cdots+h_{d+D+1}^{2}} \, .
\end{equation}
To prove it, we note that a Dirac symbol is a superposition (\ref{eq:linearDiracGeneral}) of anti-commuting $\gamma$ matrices.  Hence, it squares to the unit matrix up to the function $f^2$, so that to flatten a Dirac symbol one needs to divide it by $f$.
The corresponding determinant is
\begin{equation}
\label{ss: determinantNonFlattened}
\small
\left|\begin{array}{cccc}
\frac{h_{1}}{f} & \frac{h_{2}}{f} & \cdots & \frac{h_{d+D+1}}{f}\\
\partial_{1}\frac{h_{1}}{f} & \partial_{1}\frac{h_{2}}{f} & \cdots & \partial_{1}\frac{h_{d+D+1}}{f}\\
\vdots & \vdots & \ddots & \vdots\\
\partial_{d+D}\frac{h_{1}}{f} & \partial_{d+D}\frac{h_{2}}{f} & \cdots & \partial_{d+D}\frac{h_{d+D+1}}{f} 
\end{array}\right| \, .
\end{equation}
Since each derivative is 
\begin{equation}
\partial_{i}\frac{h_{j}}{f}=\frac{1}{f}\partial_{i}h_{j}-\frac{h_{j}}{f^{2}}\partial_{i}f \, ,
\end{equation}
and using the identity
\begin{equation}
\left|\begin{array}{cc}
a & b \\
c + e & d + f
\end{array}\right| = \left|\begin{array}{cc}
a & b \\
c & d
\end{array}\right| + \left|\begin{array}{cc}
a & b \\
e & f
\end{array}\right| \, ,
\end{equation}
the first term provides the result (\ref{eq:DiracDeterminant}), while the second term vanishes since its rows are linearly dependent.

Relation (\ref{eq:conditions}) allows to identify  $\mathbb{Z}$-topology classes for which either Chern or winding numbers may not vanish. It implies that $p+q$ and $d+D$ have the same parity, as well as $s = p -q $ and $\delta = d - D$. Hence, their difference $s - \delta$ must be even, namely odd values ensure the absence of $\mathbb{Z}$-topology. Moreover, under  $\boldsymbol{k}\rightarrow -\boldsymbol{k}$, the determinant (\ref{eq:DiracDeterminant}) is modified by an overall factor $(-1)^{d-p}$ of which $d$ variables  $\boldsymbol{k}$ for each derivative and $p$ antisymmetric variables $\boldsymbol{h}_a$. Since (\ref{eq:chern_winding_determinant}) is evaluated over an even  $\boldsymbol{k}$-domain, it vanishes for $d-p = 1 \;(\text{mod}\, 2)$. Noting that (\ref{eq:conditions}) implies $
d-p  =  \frac{\delta-s}{2}$, we infer that for $\delta-s = 2 \;(\text{mod}\, 4)$, the corresponding classes do not have $\mathbb{Z}$-topology. Finally, the only classes with topological integer numbers correspond to  
\begin{equation}
\label{eq:s_delta_topology}
\delta-s \equiv 0 \;(\text{mod}\, 4) \, .
\end{equation}
This result has been established otherwise by means of sophisticated methods \cite{Teo2010}. Here, it is a direct consequence of properties of the Jacobian (\ref{eq:DiracDeterminant}). The novelty is in the introduction of $D$ additional degrees of freedom which allow to change the number of symmetric $(\boldsymbol{h}_s)$ and antisymmetric $(\boldsymbol{h}_a)$ fields, hence modifying the scaling parameter $s-\delta$. The dimension $D$ in (\ref{eq:conditions}) appears as a way to construct $\mathbb{Z}$-topological defect fields. To implement it, we consider Dirac symbols, 
% \footnote{Such a term is sometimes called a mass term.} 
\begin{equation} \label{eq:linearDirac} 
\mathcal{H}\left(\boldsymbol{k},\boldsymbol{r}\right) =\boldsymbol{k}\cdot\boldsymbol{\gamma_a}+\boldsymbol{\phi}\left(\boldsymbol{r}\right)\cdot\boldsymbol{\gamma_s} \, 
\end{equation}
with linear momentum dependence and spatially dependent fields $\boldsymbol{\phi}\left(\boldsymbol{r}\right)$. 
Evaluating (\ref{eq:DiracDeterminant}) and   integrating over $\boldsymbol{k}$ in (\ref{eq:chern_winding_determinant}), leads to, 
\begin{equation} 
\small
\label{DiracDeterminantReduced} 
J(\boldsymbol{\phi}, D) =
%\frac{1}{\left|\boldsymbol{\phi}\right|^{D+1}}
\left|\begin{array}{ccccc}
\phi_{1} & \phi_{2} & \cdots & \phi_{D+1}\\
\partial_{1}\phi_{1} & \partial_{1}\phi_{2} & \cdots & \partial_{1}\phi_{D+1}\\
\vdots & \vdots & \ddots & \vdots\\
\partial_{D}\phi_{1} & \partial_{D}\phi_{2} & \cdots & \partial_{D}\phi_{D+1}
\end{array}\right| \, 
\end{equation}
so that $\nu_{d+D}=\nu_{D}$ and  $\delta = s$ (the first $\mathbb{Z}$ diagonal in Table \ref{tab:10-fold_way}). Relation  (\ref{eq:conditions}) implies $d = p$ and $D = q$, therefore $D$  characterises the defect fields $\boldsymbol{\phi}\left(\boldsymbol{r}\right)$ in  analogy with the defect classification in thermodynamic phase transitions \cite{Toulouse1976}. A zero-dimensional defect,  $D=q=0$ corresponds to a real valued scalar field and a domain wall, 
$D=q=1$ to a complex scalar field, $q=2$ to a vector field, etc. 

\section{Topological defects}
\label{sec:defects}
We now review examples summarized in Table \ref{tab:topological_systems}. 
\subsection{The SSH model}
At low energies, the well-known SSH model \citep{SSH1979, Heeger1988} is described by the symbol 
\begin{equation}
\label{eq:ssh_symbol}
\mathcal{H}\left(k\right) = k\sigma_x +m \, \sigma_y \, ,
\end{equation}
characterised by $d = p = 1$ and $ q = 0$. %and a winding number $1, 0$ depending on the sign of $m$. 
A position-dependent $m(\boldsymbol{r})$ does not change $q$. It corresponds to a real scalar field so that (\ref{eq:conditions}) requests $D=0$ to display a nontrivial topology, i.e. a domain wall with winding number $\pm1$ \cite{Chiu2016}.
\subsection{Graphene with a vacancy}
The Dirac symbol (\ref{eq:graphene_bloch_dirac}) of pristine monolayer graphene involves a pair $(\gamma_s,  \gamma_a) $ of Dirac matrices, so that $s=p-q =1$. %making it a member of $BDI$. %Since $\delta = d =2$,  $s \neq \delta$, so that graphene does not admit $\mathbb{Z}$ topology., 
In the absence of defect field, namely for  $\boldsymbol{\phi}\left(\boldsymbol{r}\right) = 0$,  $D \equiv 0$ and $d+D=2\neq p+q = 1$. Hence (\ref{eq:conditions}) is not fulfilled and graphene does not display $\mathbb{Z}$-topology. Adding a vacancy, the corresponding symbol  (\ref{eq:H_graphene}) fulfils  (\ref{eq:linearDirac}). The vacancy field $\boldsymbol{\phi} (\boldsymbol{r})$  involves two symmetric $\gamma_s$, i.e. $p=2,q=1,d=2$ and 
 $s=1$. Fulfilling (\ref{eq:conditions}) requires $D=1$, so that $\boldsymbol{\phi} (\boldsymbol{r})$ is a complex valued field. All together, graphene with a vacancy belongs to the $BDI$ class with $\delta = d - D =1$, and a topological number calculated from  (\ref{eq:chern_winding_determinant}) and (\ref{DiracDeterminantReduced}), 
\begin{equation} \label{eq:windingGraphene} %\begin{aligned}
\nu_{3} =\frac{1}{2\pi}\int dr \, \frac{1}{\phi_{1}^{2}+\phi_{2}^{2}}\left|\begin{array}{cc}
\phi_{1} & \phi_{2}\\
\partial_{r}\phi_{1} & \partial_{r}\phi_{2}
\end{array}\right| 
 %& =\frac{1}{2\pi}\int dr\partial_{r}\Delta\\
 % =\frac{1}{2\pi}\int d\Delta 
 =\mp \int \frac{d\theta}{2\pi}=\mp 1 
%\end{aligned} 
\end{equation}
which is the winding of the phase %$\pm\theta$ 
of $\boldsymbol{\phi} (\boldsymbol{r})$,   
 encircling the vacancy, where $-1 (+1)$ respectively corresponds to a vacancy on sublattice $A (B)$. 
%\begin{equation} \label{eq:windingResult} 
%\nu_{3}=\pm\frac{1}{2\pi}\int d\theta=\pm 1
%\end{equation}
%Moreover, a vacancy induces an interaction between the two valleys $(K,K')$ decoupled in graphene.
\subsection{Graphene with an adatom}
An adatom, another kind of spatial defect in graphene, shares several features with a vacancy, e.g. both have zero energy modes spatially located on the impurity, but they have distinct topological properties \citep{Dutreix2016,Nanda2013}. Unlike vacancies, creating an adatom  involves a new  energy scale which breaks particle-hole and chiral symmetries, suggesting it belongs to $AI$ rather than $BDI$ class. These features can be understood by adding to  (\ref{eq:graphene_bloch_dirac}) a mass term,
\begin{equation} \label{eq:H_graphene_adatom}
\mathcal{H}_A (\boldsymbol{k}, \boldsymbol{r}) =h_1(\boldsymbol{k})\, \sigma_{x}+h_2(\boldsymbol{k}) \, \sigma_{y} + m(\boldsymbol{r}) \, \sigma_{z}
\end{equation}
where $h_1-ih_2 = 1 + e^{-i\boldsymbol{k}\cdot \boldsymbol{a}_1} + e^{-i\boldsymbol{k}\cdot \boldsymbol{a}_2}$.
%This Dirac symbol has $d=2, D=1$. 
Note that we only consider a single valley since the mass term does not couple the two valleys of graphene. Since $h_2(-k) = -h_2(k)$ and $h_1(-k) = h_1(k)$,
the three Dirac matrices fulfill  $p=1$ and $ q=1$, so that $s=0$ which indeed corresponds to the $AI$ class. Relation  (\ref{eq:conditions}) is satisfied for $D=0$, implying $\delta=d-D=2 \neq s $ so that graphene with an adatom does not display $\mathbb{Z}$-topology.
\subsection{Graphene with vacancy and adatom}
The coexistence of a vacancy and an adatom in graphene raises the  experimentally relevant question of the topological prominence of a given defect field. 
%The previous section is worry some as it suggests that a single adatom breaks the chiral symmetry of the Hamiltonian and leads to trivial topology, so even if we had a graphene with vacancy, a single adatom will make it lose its topological features.
The symbol of graphene with a vacancy and an adatom at distinct locations, is of the type (\ref{eq:linearDirac}),
\begin{equation}
\label{eq:symbol_G_V_A}
\mathcal{H}_{A+V}\left(\boldsymbol{k},\boldsymbol{r},
\boldsymbol{r'}\right)  = \mathcal{H}_V \left(\boldsymbol{k},\boldsymbol{r}\right) +  m(\boldsymbol{r}^\prime)\sigma_z\otimes\boldsymbol{1}. 
\end{equation}
It involves two fields, one complex and one scalar, so that   
%The vacancy and the impurity are each encircled by a sphere of dimension 1 and so we have $D=2$. 
%This symbol has 
$p=2$, $q=2$ and $s=0$ ($AI$ class). Relation (\ref{eq:conditions}) requests $D=2$, namely an overall vector field for the combined defect (vacancy and adatom). Therefore, while the adatom breaks chiral symmetry, $s = \delta =0$, the vacancy + adatom system still displays $\mathbb{Z}$-topology. Since $d+D$ is even, from (\ref{DiracDeterminantReduced}), the corresponding invariant is 
%, a result showing that the $\mathbb{Z}$ topology induced by a vacancy is unaffected by the presence of an adatom
the second Chern number
\begin{equation} 
\footnotesize
\label{eq:ChernGraphene} 
\mathcal{C}_{2} = \frac{1}{4\pi} \int dr dr^\prime \frac{\partial_{r^\prime}m}{(\phi_{1}^{2}+\phi_{2}^{2} + m^2)^{3/2}}\left|\begin{array}{cc}
\phi_{1} & \phi_{2}\\
\partial_{r}\phi_{1} & \partial_{r}\phi_{2}
\end{array}\right|  = \mp1.
\end{equation}
\subsection{Graphene with a Kekule distortion}
Another defect field which came recently under scrutiny is the valley coupling Kekule distortion obtained by shifting the hopping term in (\ref{eq:graphene_NN_tb}) into $t_{\boldsymbol{r}, i}=t+\delta t_{\boldsymbol{r}, i}$ \cite{Hou2007,Jackiw2007, Gomes2012, Qu2022, Bao2021} with
\begin{equation}
\delta t_{\boldsymbol{r},i}=\Delta\left(\boldsymbol{r}\right)\exp{\left( i\boldsymbol{K}\cdot\boldsymbol{\delta}_{i} + i\Delta\boldsymbol{K}\cdot\boldsymbol{r} \right)}.
\end{equation} 
The corresponding symbol is  (\ref{eq:H_graphene}) with $\boldsymbol{\phi} (\boldsymbol{r})  \equiv \Delta(\textbf{r})=\Delta_0\left(r\right)e^{i(\alpha +n\theta)}$ and constant $\alpha$. The Kekule model and the vacancy are thus equivalent topological defects characterised by $D=1$ (see Table \ref{tab:topological_systems}) and $\mathbb{Z}$-topology but with $\nu_3 = n$ instead of (\ref{eq:windingGraphene}). The Kekule distortion exhibits topological fractional charge and so does a vacancy \cite{Ovdat2017,Ovdat2020}. %However, these two models have very distinct spatial features. 
\begin{table}[ht]

\centering

\begin{tabular}{| c | c | c | c | c | c | c | c |}
    \hline
     & $p$ & $d$ & $q$ & $D$ & $s$ & $\delta$ & Topology \\ 
    \hline
    
    \text{Graphene} & 1 & 2 & 0 & 0 & 1 & 2 & No \\ 
    SSH & 1 & 1 & 0 & 0 & 1 & 1 & Yes \\ 
    $\text{Vacancy}$ & 2 & 2 & 1 & 1 & 1 & 1 & $\text{Yes}$ \\
    Kekule & 2 & 2 & 1 & 1 & 1 & 1 & Yes
    \\
    Adatom & 1 & 2 & 1 & 0 & 0 & 2 & No \\
    V + A & 2 & 2 & 2 & 2 & 0 & 0 & Yes \\
    \hline
    
\end{tabular}
\caption{\label{tab:topological_systems} Defects fields in graphene. To display a $\mathbb{Z}$-topology (last column), the dimension $D$ must fulfill $D =p+q-d$. V+A stands for vacancy + adatom.}
\end{table}

\section{The Atiyah-Singer index theorem}
\label{sec:index thorem}
\subsection{Analytical and topological indices}
The existence of $\mathbb{Z}$ topological defects is related to the appearance of zero energy modes in the spectrum of the Hamiltonian \citep{Ryu2002, Brouwer2002, Ganeshan2013}. For a vacancy, this relation allows to calculate $\nu_{2m-1}$ in (\ref{eq:windingDefinition}) using the Atiyah-Singer index theorem \citep{Atiyah1963, Atiyah1968} relating properties of a Hamiltonian and its symbol. This theorem states that the analytical index, counting zero modes of an  elliptic differential operator defined on compact manifolds, is equal to the topological index calculated from its symbol \cite[Chapter~ 24-27]{yankowsky2013}. Here, the elliptic differential operator is the Hamiltonian and the compact manifold is the Brillouin zone in the presence of translation symmetry or $S^d$ otherwise. For a chiral Hamiltonian, 
\begin{equation}
\label{eq:chiralHamiltonian}
H= \begin{pmatrix}
        0 & \cal{Q} \\
        \cal{Q}^{\dagger} & 0 
    \end{pmatrix},
\end{equation}
the analytical index is 
\begin{equation}
\label{eq:analyticalIndex}
\text{Index} \, H \equiv \text{dim}\, \text{Ker} \, \cal{Q}  -  \text{dim}\, \text{Ker} \, \cal{Q}^{\dagger}
\end{equation}
and the topological index is the  
 winding (\ref{eq:windingDefinition}).
For a vacancy in sublattice B (A), the Hamiltonian $H_V$ in (\ref{eq:graphene_NN_tb_vacancy}) is chiral, elliptic and hosts a single zero mode located on sublattice A (B), hence its index is $\pm1$ as given by (\ref{eq:windingGraphene}), and the Atiyah-Singer theorem becomes,
\begin{equation}
\label{eq:AnalyticalIndex}
\text{Index} \, H_V \equiv \text{dim}\, \text{Ker} \, {\cal{Q}}  -  \text{dim}\, \text{Ker} \, {\cal{Q}}^{\dagger} = \nu_3 \, .
\end{equation}
The index is still defined for a non-chiral Hamiltonian, e.g, 
\begin{equation}
\label{eq:chiralHamiltonianMass}
H=\begin{psmallmatrix}
M & \cal{Q} \\
\cal{Q}^{\dagger} & -M 
\end{psmallmatrix}
\end{equation}
where a mass term $M$ shifts the energy spectrum. In that specific case, both the eigenstates and  $\text{Index} \, H$ remain unchanged but  zero modes of $\cal{Q}$ and $ \cal{Q}^\dagger$ belong to the spectrum only for a chiral Hamiltonian. 
For a space-dependent mass term, this illustrates what happens when adding an adatom to  graphene with a vacancy. The adatom breaks chiral symmetry but preserves topology, viz. the index theorem (\ref{eq:AnalyticalIndex}) holds which relates the Chern number $\mathcal{C}_{2}$ in (\ref{eq:chernDefinition})
 and the analytical index. 
\subsection{Bulk-Edge correspondence}
Bulk-edge correspondence relates bulk topological numbers to edge states. For a vacancy with winding number $\pm1$, a single edge state is expected whose location depends on boundary conditions (Fig.\ref{fig:ZM_example_one_vacancy_no_disorder_N5_armChair_zigzag}).
%, either on the vacancy (a), along the lattice edge (c), or both (b).
For any type of boundary conditions, the zero mode is located either on the edge of the lattice, on the vacancy location, or both. This suggests considering a vacancy  as an additional edge of the lattice.
\subsection{Generalization to finitely many vacancies}
For any number $(V_A,V_B)$ of vacancies, a generalisation, 
\begin{equation}
\text{Index}\,  H_V = V_A-V_B = \nu_3 \, ,
\label{eq:ASIndex}
\end{equation}
of the index theorem (\ref{eq:AnalyticalIndex}), allows to count edge states as shown in Fig.\ref{fig:ZM_example_one_vacancy_no_disorder_N5_armChair_zigzag} (e) for three vacancies, two on sublattice $A$ and one on $B$ so that $\nu_3 = 1$, hence corresponding to a single edge state \cite{Ovdat2020}. For a dimer and generally for $V_A = V_B$, $\nu_3 = 0$, and no edge state is expected despite the presence of vacancies. 
\subsection{Disorder}
\begin{figure}[ht]
    \centering
    \begin{tabular}{cc}
      \includegraphics[width=0.22\textwidth]{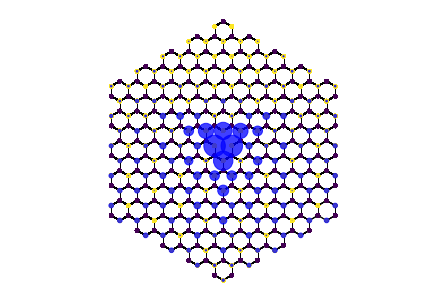}
      &
      \includegraphics[width=0.22\textwidth]{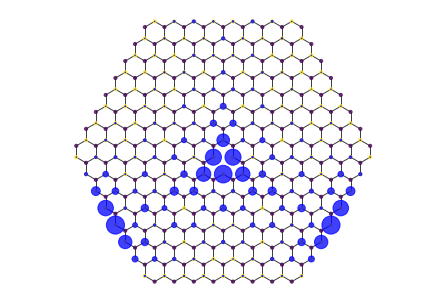}
      
      \\
      \small (a) & \small (b) 
      \\

      \includegraphics[width=0.22\textwidth]{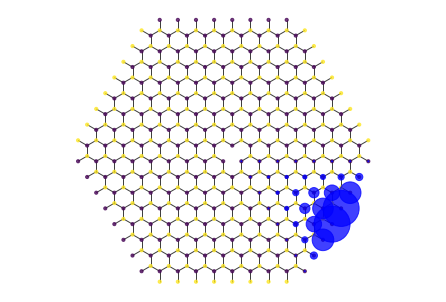}
      &

      \includegraphics[width=0.22\textwidth]{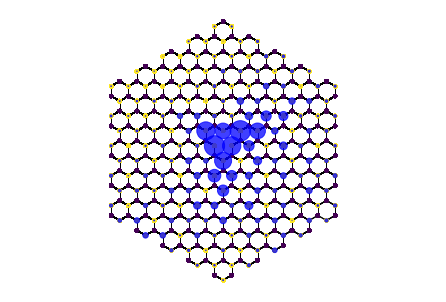}
      \\
        
    %  \includegraphics[width=0.25\textwidth]{ZM_example_two_vacancy_in_no_disorder_part_1.png}
    %  \\
    %  \small (c) & \small (d)
    %  \\
   %   \includegraphics[width=0.25\textwidth]{ZM_example_two_vacancy_in_no_disorder_part_2.png}
    %  &
      
    \small (c) & \small (d)
      \\
    \includegraphics[width=0.22\textwidth]{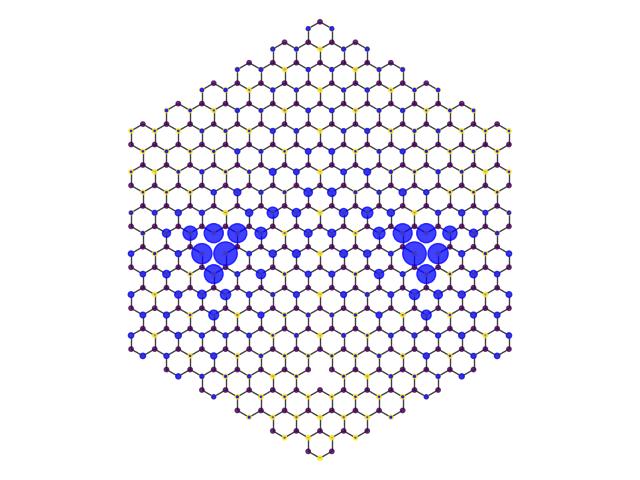}
     \\
     \small (e)
	\end{tabular}
	\caption{Zero energy edge state from exact diagonalization of (\ref{eq:graphene_NN_tb_vacancy}). (a), (b) and (c) correspond to  armchair, zigzag and bearded boundary conditions, (d) in the presence of disorder. 
 (e) For three vacancies, two on the $B$ sublattice and one on $A$, a single edge state is expected from (\ref{eq:ASIndex})  located on the majority $B$ sublattice.
 }
	\label{fig:ZM_example_one_vacancy_no_disorder_N5_armChair_zigzag}
\end{figure}

An important property of topological edge states is their robustness against perturbations e.g. disorder. Introducing a random hopping term in (\ref{eq:graphene_NN_tb_vacancy}), still preserves the topological zero energy edge state and its winding $\nu_3 = \mp 1$ as displayed in Fig. \ref{fig:ZM_example_one_vacancy_no_disorder_N5_armChair_zigzag} (d).  
%\subsection{Summary}

\section{Applicability of index theorems}
\label{sec:no_AS}
The Atiyah-Singer (AS) index theorem \citep{Atiyah1963, Atiyah1968} is often invoked to identify zero energy states of a Hamiltonian as an edge (or boundary) mode and to relate them to topological (bulk) numbers. This procedure relating edge physics to bulk topological features constitutes the so-called bulk-edge correspondence. It is sometimes invoked improperly and it is the purpose of this section to discuss its relevance in more general terms using interesting and relevant  limiting cases.  The analytical index defined in (\ref{eq:analyticalIndex}), 
\begin{equation}
\text{Index} \, H = \text{dim Ker} \, \mathcal{Q} - \text{dim Ker} \, \mathcal{Q}^\dagger,
\end{equation}
characterises chiral Hamiltonians of the form (\ref{eq:chiralHamiltonian}). An important content of the AS theorem states that the number of  zero modes counted by the analytical index is a topological invariant which can be expressed and related to the topological index of the symbol given either by a Chern or winding number.
As appealing as it seems, it is important to pay attention to the strict applicability conditions of this  theorem. It applies only to Fredholm operators or to elliptic differential operators defined on compact domains (hence a Fredholm operator). This class includes  differential operators with invertible symbol namely having a gap, except for $\boldsymbol{k}=0$ but with a finite multiplicity. Hence both zero modes counting spaces $\text{Ker} \, \mathcal{Q}$  and $\text{Ker} \, \mathcal{Q}^\dagger$ are finite dimensional. For example, the Laplacian is an elliptic differential operator whose symbol $k^2$, vanishes only at $\boldsymbol{k} = 0$.
The Hamiltonian of graphene with a vacancy is also an elliptic differential operator as its symbol has a gap as given in (\ref{eq:H_graphene}) and displayed in Fig.\ref{fig:Hamiltonian_Symbol_spectrum}.

 Consider now two other systems, the nearest neighbours tight-binding model on a square lattice, essentially a discrete version of the Laplacian operator, and the Lieb lattice \cite{lieb1989}, both of which having zero energy modes and a chiral Hamiltonian of the form (\ref{eq:chiralHamiltonian}). In both cases, the AS index theorem does not apply since the corresponding Hamiltonians are not elliptic differential operators. A square lattice displays apparent yet fictitious chiral symmetry. Although the lattice splits into two sublattices, this partition is unnecessary and misleading, since topological properties require exhausting all unitary symmetries.
The symbol of the Hamiltonian of a square lattice, i.e. its Bloch Hamiltonian, is given by the function,
\begin{equation} \label{SL}
\mathcal{H}_{SL}=-2t\left(\cos{k_xa} + \cos{k_ya} \right).
\end{equation}
This symbol vanishes along the lines $k_x\pm k_y=\pi$ and $k_x\pm k_y = -\pi$ so that the square lattice tight binding Hamiltonian is not elliptic. The low energy limit of the symbol (i.e the Bloch Hamiltonian) of the Lieb lattice \cite{lieb1989} Hamiltonian is given by
\begin{equation} \label{LL}
\mathcal{H}_{L}=-ta\boldsymbol{k}\cdot \boldsymbol{L}
\end{equation} where
\begin{equation}
L_x = \begin{pmatrix}
0 & -i & 0 \\ 
i & 0 & 0 \\ 
0 & 0 & 0 \\ 
\end{pmatrix}, \; L_y = \begin{pmatrix}
0 & 0 & -i \\ 
0 & 0 & 0 \\ 
i & 0 & 0 \\ 
\end{pmatrix}.
\end{equation}
A diagonalisation of $\mathcal{H}_{L}$ shows that this symbol has a single Dirac cone and  a flat band of zero energy modes, hence it is also not elliptic.

Considering the topology of these two systems, we immediately check that the square lattice is not topological. It belongs to the $AI$ class (See Table \ref{tab:10-fold_way}) since the symbol (Bloch Hamiltonian) (\ref{SL}) has time reversal symmetry $T=K$ but no particle hole symmetry $\left( s=0 \right)$ and it has $\delta = d-D = 2-0 =2$, hence void of topological invariant numbers. Adding a vacancy to this system will still give trivial topology since $ \delta = d-D = 2-1=1$.
The Lieb lattice symbol (\ref{LL}) does not have gap and it cannot be expressed by means of Dirac matrices of a Clifford algebra, so that it  does not belong to the tenfold classification. Yet, it may display topological features accessible from a version of the Atiyah Singer index theorem different from (\ref{eq:AnalyticalIndex}) \cite{Venaille2021}. This possibility is ruled out because  index theorems apply to elliptic operators whose spectrum vanishes  for finitely many eigenvalues \cite{yankowsky2013}, in contradiction with the existence of a zero energy flat band.

\begin{figure}[ht]
    \centering
    \begin{tabular}{cc}
      \includegraphics[width=0.25\textwidth]{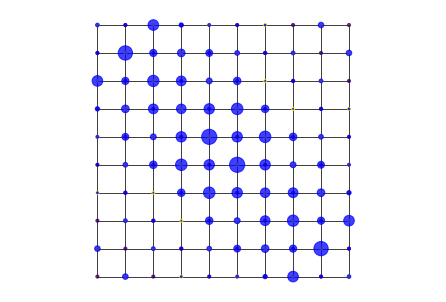}
      &
      \includegraphics[width=0.25\textwidth]{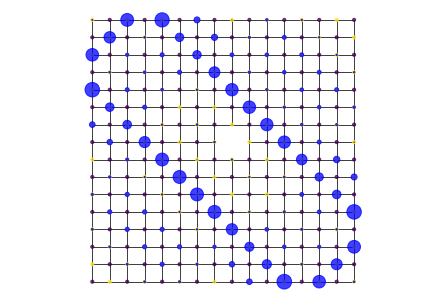}
      \\
      
       \small (a) & \small (b) \\
       \includegraphics[width=0.25\textwidth]{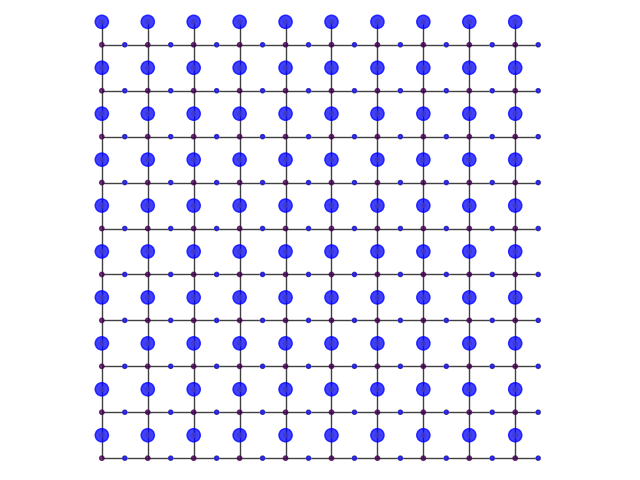}
      \\
     
      \small (c)

	\end{tabular}
	\caption{Wavefunctions (squared) associated to zero modes (in blue) obtained by exact diagonalization of nearest-neighbour tight binding Hamiltonians. (a) A zero mode on a square lattice with Dirichlet boundary conditions. (b) A zero mode on a square lattice with Dirichlet boundary conditions and a single vacancy. (c) A zero mode on a Lieb lattice with periodic boundary conditions. These zero modes are not topological  and they are not edge states.}
	\label{fig:ZM_SL_Lieb_lattice}
\end{figure}

However, both Hamiltonians have zero modes displayed in Fig. \ref{fig:ZM_SL_Lieb_lattice}. These zero modes are not of topological origin as the index theorem is not valid in these cases. Said otherwise, these zero modes are not edge states, they are not localized on the edges of the lattice, nor on the vacancy location, and they are not protected against perturbations, e.g. they are sensitive to disorder.

\section{Conclusion}

We have introduced defect fields characterised by a dimension $D$ and symbol operators generalising Bloch Hamiltonians, to tailor topological features of materials describable by the tenfold classification. This approach should prove useful to use appropriate defects or textures to create quantum materials with topological features on demand, namely to allow navigating between different entries of the tenfold classification. 

More specifically, we have considered valley-coupling defect fields $\boldsymbol{\phi}\left(\boldsymbol{r}\right)$ in graphene or alike materials. The existence of invariant numbers (Chern or winding) translate into edge states and their relation to powerful Index theorems has been emphasized. %often assimilated to zero modes. 
Defect fields have been considered e.g. vacancy and adatom which, despite  common features, are not topologically akin. We have discussed examples where zero energy modes exist which are not related to any topological features, hence care is required when 
extrapolating results obtained in the framework of the tenfold classification. 

\label{sec:conclusion}

\begin{acknowledgments}
This work was supported by the Israel Science Foundation Grant No.~772/21 and by the Pazy Foundation.
\end{acknowledgments}.
\appendix

\section{Symbol of graphene with a vacancy}
\label{sec:app_a}

In this Appendix, we establish expression (\ref{eq:H_graphene}) for the symbol of graphene with a vacancy. Since topological obstructions if at all show up at any energy scale, it is sufficient to establish the expression of the symbol in the low energy limit.
The low energy contribution of a vacancy in sublattice $A$ to the tight binding Hamiltonian of graphene is given, in appropriate units, by the $4\times4$ matrix, %\cite{Abulafia2023},
\begin{equation}
\footnotesize
\label{eq:vacancy_potential}
V \left( \boldsymbol{k},\boldsymbol{r} \right)=\left(\begin{array}{cccc}
0 & 0 & \delta\left(\boldsymbol{r}\right)Q & -\delta\left(\boldsymbol{r}\right)Q^{\dagger}\\
0 & 0 & \delta\left(\boldsymbol{r}\right)Q & -\delta\left(\boldsymbol{r}\right)Q^{\dagger}\\
Q^{\dagger}\delta\left(\boldsymbol{r}\right) & Q^{\dagger}\delta\left(\boldsymbol{r}\right) & 0 & 0\\
-Q\delta\left(\boldsymbol{r}\right) & -Q\delta\left(\boldsymbol{r}\right) & 0 & 0
\end{array}\right),
\end{equation}
where the wave functions are written in the sublattice basis $\psi_{\boldsymbol{r}}=\left(\begin{array}{cccc}
a_{\boldsymbol{r}}^{K} & a_{\boldsymbol{r}}^{K'} & b_{\boldsymbol{r}}^{K} & b_{\boldsymbol{r}}^{K'}\end{array}\right)^{T}$ and $Q^{\dagger}=k_{x}+ik_{y},\,Q=k_{x}-ik_{y}$. Note that  $\boldsymbol{r},\boldsymbol{k}$ are (non commuting) operators.

To find the symbol of $V\left(\boldsymbol{r},\boldsymbol{k} \right)$, we need to calculate its Weyl transform defined in (\ref{eq:Weyl_transform}). While this calculation is feasible, there is a simpler way to achieve the result. A property of the inverse Weyl transform, i.e. the Wigner transform, is that it transforms a classical Hamiltonian into its Weyl ordered quantum counterpart. The Weyl ordered Hamiltonian is obtained by complete symmetrization, e.g., $kr\rightarrow \frac{\hat{k}\hat{r} +\hat{r}\hat{k}}{2}$. We use it as follows. Each component in  (\ref{eq:vacancy_potential}) is written as a sum of a symmetric and an anti-symmetric part with respect to $\boldsymbol{k}$ and $\boldsymbol{r}$, e.g.,
\begin{equation}
\delta\left(\boldsymbol{r}\right)Q = \frac{\left\{\delta\left(\boldsymbol{r}\right),Q\right\}}{2} +\frac{\left[\delta\left(\boldsymbol{r}\right),Q\right]}{2},
\label{S2}
\end{equation}
where $\left\{,\right\}$ and $\left[,\right]$ are respectively anti-commutation and commutation relations.
Since $Q$ is linear in $k_x,k_y$ the second term in (\ref{S2}) is given by $\frac{-i\left(\partial_x-i\partial_y\right)\delta\left(\boldsymbol{r}\right)}{2}$, i.e., it is a function of $\boldsymbol{r}$ only and not of $\boldsymbol{k}$, hence its symbol is given by replacing the operator $\hat{\boldsymbol r}$ by the parameter $\boldsymbol{r}$. The first term is Weyl ordered, so its Weyl symbol is given by the function $\delta\left(\boldsymbol{r}\right)Q$, where  $\left( \boldsymbol{r},\boldsymbol{k} \right)$ are parameters. Overall, using the standard notation $W\left[ \cdots \right]$ for the symbol of an operator, we obtain, 
\begin{equation}
\label{eq:symbol_two_terms}
W\left[\delta\left(\boldsymbol{r}\right)Q\right] = \delta\left(\boldsymbol{r}\right)Q - \frac{i\left(\partial_x-i\partial_y\right)\delta\left(\boldsymbol{r}\right)}{2}.
\end{equation}
Note that on the right hand side of (\ref{eq:symbol_two_terms}) $\boldsymbol{r},\boldsymbol{k}$ are parameters, and will be so henceforth.
Comparing the magnitude of the two terms in (\ref{eq:symbol_two_terms}), we have
\begin{equation}
\label{eq:size_comparison}
\begin{aligned}
& \delta\left(\boldsymbol{r}\right)Q - \frac{i\left(\partial_x-i\partial_y\right)\delta\left(\boldsymbol{r}\right)}{2} \\ & \sim \frac{1}{2a} \delta\left(\boldsymbol{r}\right)\left[\left(2k_xa-i\right)-i\left(2k_ya-i\right)\right],
\end{aligned}
\end{equation}
where $a$ is the lattice constant. In the low energy limit  $ka\ll1$, the dominant term in (\ref{eq:symbol_two_terms}) is the second one that we keep. 
Moreover, using that $\delta\left(\boldsymbol{r}\right)$ is a function of $r$ only and not of $\theta$, leads to 
\begin{equation}
\label{eq:delta_derivative}
\begin{aligned}
-i\left(\partial_x\pm i\partial_y\right)\delta\left(\boldsymbol{r}\right) & = e^{\pm i\theta}\left(-i\partial_r\pm\frac{1}{r}\partial_\theta\right)\delta\left(\boldsymbol{r}\right) \\ & = -ie^{\pm i\theta}\partial_r\delta\left(\boldsymbol{r}\right) \, .
\end{aligned}
\end{equation}
Overall, the symbol of the Hamiltonian of graphene with an $A$ vacancy is the sum $W\left[H_0 + V \left( \boldsymbol{r},\boldsymbol{k} \right) \right]$, where the symbol of the unperturbed Hamiltonian $H_0$ in (\ref{eq:graphene_NN_tb}), is 
\begin{equation}
\label{eq:symbol_H_0}
W\left[H_0\right]=\left(\begin{array}{cccc}
0 & 0 & Q & 0\\
0 & 0 & 0 & -Q^{\dagger}\\
Q^{\dagger} & 0 & 0 & 0\\
0 & -Q & 0 & 0
\end{array}\right),
\end{equation}
namely the Bloch Hamiltonian (\ref{eq:graphene_bloch_dirac}). The total symbol involves two type of terms, those which are function of $\boldsymbol{r}$ and those which are functions of $\boldsymbol{k}$ and $\boldsymbol{r}$. For the latter case, the $\boldsymbol{r}$-dependant part can be discarded. To see it, consider e.g, the term 
\begin{equation}
\label{eq:symbol_not_highest_order}
 Q -\frac{ie^{- i\theta}\partial_r\delta\left(\boldsymbol{r}\right)}{2} \, .   
\end{equation} 
The only contribution needed to extract relevant topology is the principle symbol of the Hamiltonian, obtained by considering only the highest order in $\boldsymbol{k}$ (highest order derivatives in the Hamiltonian) \cite{yankowsky2013}. It is the first term in (\ref{eq:symbol_not_highest_order}), hence we can discard the second term.
Notice that for the anti-diagonal terms, this procedure does not apply since for them the principle symbol is given by $-ie^{\pm i\theta}\partial_r\delta\left(\boldsymbol{r}\right)$, these terms being independent of $\boldsymbol{k}$.

Finally, we wish to emphasize that $\delta\left(\boldsymbol{r}\right)$ is an approximation for the potential created by a vacancy. Generally, it can be replaced by some spatially localized function, whose exact shape is not important. Hence we take $\frac{\partial_r\delta\left(r\right)}{2} \rightarrow \phi\left(\boldsymbol{r}\right)$. Together with the replacement $\theta\rightarrow \theta + \frac{\pi}{2}$ so as to absorb the $-i$ prefactor, we obtain,
\begin{equation}
\footnotesize
\label{eq:symbol} \mathcal{H}_V (\boldsymbol{k}, \boldsymbol{r}) =\left(\begin{array}{cccc}
0 & 0 & Q & e^{i\theta}\phi\left(r\right)\\
0 & 0 & e^{-i\theta}\phi\left(r\right) & -Q^{\dagger}\\
Q^{\dagger} & e^{i\theta}\phi\left(r\right) & 0 & 0\\
e^{-i\theta}\phi\left(r\right) & -Q & 0 & 0
\end{array}\right)
\end{equation}
for the symbol of $H_V$, restoring the notations used in (\ref{eq:H_graphene}). This symbol when expressed in terms of Dirac matrices, 
\begin{equation} \begin{aligned}
\mathcal{H}_V (\boldsymbol{k}, \boldsymbol{r}) & =k_{x}\, \sigma_{x}\otimes\tau_{z}+k_{y} \, \sigma_{y}\otimes \mathbf{1} \\ & + {\phi}_{1} \left(\boldsymbol{r} \right)\sigma_{x}\otimes\tau_{x}+ \phi_{2}\left(\boldsymbol{r} \right)\sigma_{x}\otimes\tau_{y}\\
%  & =\left(\begin{array}{cccc}
% 0 & 0 & k_{x}-ik_{y} & \phi e^{-i\Delta}\\
% 0 & 0 & \phi e^{i\Delta} & k_{x}+ik_{y}\\
% k_{x}+ik_{y} & \phi e^{-i\Delta} & 0 & 0\\
% \phi e^{i\Delta} & k_{x}-ik_{y} & 0 & 0
% \end{array}\right)
\end{aligned} 
\end{equation}
with 
$ \phi_{1}\left(\boldsymbol{r} \right)+i \phi_{2}\left(\boldsymbol{r} \right)= \phi\left(r \right) \, e^{-i\theta}$ is the relation (\ref{eq:H_graphene}).

\section{Construction of the 10-fold way with the symbol}
\label{sec:app_b}
In this section we follow \cite{Teo2010}. We work with Dirac symbols given by (\ref{eq:Diracdefnition}), of the form
\begin{equation}
\small
\mathcal{H}\left(\boldsymbol{k},\boldsymbol{r}\right)=\boldsymbol{h}_s\left(\boldsymbol{k},\boldsymbol{r}\right)\cdot\boldsymbol{\gamma}_s+\boldsymbol{h}_a\left(\boldsymbol{k},\boldsymbol{r}\right)\cdot\boldsymbol{\gamma}_a \, ,
\end{equation}
where \begin{equation} 
\begin{aligned}
\boldsymbol{h}_s\left(\boldsymbol{k},\boldsymbol{r}\right) & =\boldsymbol{h}_s\left(-\boldsymbol{k},\boldsymbol{r}\right) \\
\boldsymbol{h}_a\left(\boldsymbol{k},\boldsymbol{r}\right) & =-\boldsymbol{h}_a\left(-\boldsymbol{k},\boldsymbol{r}\right) .
\end{aligned} 
\end{equation}
We denote by $p$ (resp. $q+1$) the number of $\gamma_a$ (resp. $\gamma_s$). Momentum and position variables $ \boldsymbol{k} $ and $ \boldsymbol{r} $, are respectively defined by the $d$ components $(k_1, \dots, k_d)$ on the $S^d$ sphere
and the $D$ spatial components $ (r_1, \dots, r_D)$ of the defect on $S^D$.

The purpose of this Appendix is to relate $\left(p,q \right)$ to the symmetry class of a symbol indexed by $s$, namely to show that $s=p-q$. To that aim, the  transition from one symmetry class to another is considered and then applied to Dirac symbols.

Starting from symbol in a $s$ symmetry class, we show that by adding or removing one position or momentum coordinate, a
new symbol can be constructed which belongs to another symmetry 
class. % according to the Bott clock.
It is easy to see that there are exactly two types of mappings which either add or remove symmetries, namely increasing or decreasing $s$ by one.
We start from symmetry removing mappings.
Consider a chiral symmetric symbol $\mathcal{H}_{c}$, namely anticommuting with the chirality operator $C$ (\ref{eq:symmetriesHamiltonian}), and define
the non chiral symbol 
\begin{equation} \label{eq:H_nc}
\mathcal{H}_{nc} \equiv \cos\theta \, \mathcal{H}_{c}+\sin\theta \, C
\end{equation}
where $\theta$ is position $\left( \boldsymbol{r} \right)$ or momentum $\left( \boldsymbol{k} \right)$-dependent coordinate, i.e. either D or d increases by 1. The second term
in $\mathcal{H}_{nc}$ breaks chiral symmetry and consequently (for real classes) breaks either time reversal $T$ or particle hole $P$ symmetry. To see which one is broken, we assume $\left[T,P\right] = 0$, so to rewrite the chirality operator as $C = i^{\frac{s-1}{2}}TP$ which squares to $1$ for all odd values of $s$. Both $T$ and $P$ commute with $C$ for $s = 1 \;\text{mod}\,4$ and anti-commute with $C$ for $s=3 \;\text{mod}\,4$.
From the symmetry requirements (\ref{eq:symmetriesHamiltonian}),
time reversal symmetry
is broken for $\theta$ being a momentum coordinate and $s=1 \;\text{mod}\,4$ while
particle-hole symmetry is broken for $s=3 \;\text{mod}\,4$, and $s$ increases by 1, and conversely if $\theta$
is a position dependent coordinate, $s$ decreases by 1. Hence when $s$ is odd, 
\begin{equation} \label{eq:K_resultOne} \begin{aligned}
\mathcal{H}_{c}\left(s,d,D\right) & \rightarrow \mathcal{H}_{nc}\left(s+1,d+1,D\right)\\
\mathcal{H}_{c}\left(s,d,D\right) & \rightarrow \mathcal{H}_{nc}\left(s-1,d,D+1\right) \, .
\end{aligned} 
\end{equation}

Consider now a non chiral symbol and build a chiral one by choosing, 
\begin{equation} \label{eq:H_c}
\mathcal{H}_{c}=\cos\theta \, \mathcal{H}_{nc}\otimes\sigma_{z}+\sin\theta \, \unit{\otimes}\sigma_{a}
\end{equation}
where $a=x$ or $a=y$. This symbol has chiral symmetry since
it anti commutes with $\unit \otimes i\sigma_{z} \sigma_{a}$.
The aim is now to choose $a$ such that the new symbol preserves
the original symmetry. This is also dependent on whether $\theta$
is momentum or position coordinate.
For example if $\theta$ is a momentum (position) coordinate and $\mathcal{H}_{nc}$
has time reversal symmetry, we require $a=y\left(x\right)$. Now $\mathcal{H}_{c}$
has the additional symmetry $P=\sigma_{x}T\left(i\sigma_{y}T\right)$
that satisfies $P^{2}=T^{2}\left(-T^{2}\right)$, implying $s\rightarrow s+1\left(s-1\right)$.
A similar argument holds when $\mathcal{H}_{nc}$ possess  particle hole symmetry instead.
Overall, we have thus showed that 
\begin{equation} \label{eq:K_resultTwo} \begin{aligned}
\mathcal{H}_{nc}\left(s,d,D\right) & \rightarrow \mathcal{H}_{c}\left(s+1,d+1,D\right)\\
\mathcal{H}_{nc}\left(s,d,D\right) & \rightarrow \mathcal{H}_{c}\left(s-1,d,D+1\right).
\end{aligned} 
\end{equation}
The mappings
(\ref{eq:K_resultOne}, \ref{eq:K_resultTwo}) imply that for all $s$ we have both 
\begin{equation} \label{eq:K_result} \begin{aligned}
\mathcal{H}\left(s,d,D\right) & \rightarrow \mathcal{H}\left(s+1,d+1,D\right)\\
\mathcal{H}\left(s,d,D\right) & \rightarrow \mathcal{H}\left(s-1,d,D+1\right) \, ,
\end{aligned} 
\end{equation}
namely that there is an homomorphism between symbols in class $s$ and symbols in class $s+1$ or $s-1$ and in one dimension higher, either $d$ or $D$. To show that these classes are isomorphic, an inverse map is required. The inverse map is more involved and we direct the reader to \cite{Teo2010} for the thorough proof.
By combining the mappings (\ref{eq:K_result}), the symbols  $\mathcal{H}\left(s,d,D\right)$ and  $\mathcal{H}\left(s,d+1,D+1\right)$ become isomorphic, which implies that $\mathcal{H}\left(s,d,D\right) = \mathcal{H}\left(s,\delta\right)$ where $\delta = d-D$. The mappings (\ref{eq:K_result}) can now be simplified to 
\begin{equation} \label{eq:K_resultFinal} 
\mathcal{H}\left(s,\delta\right) = \mathcal{H}\left(s+1,\delta+1\right),
\end{equation}
which finally proves the scaling
\begin{equation} \label{eq:scaling} 
\mathcal{H}\left(s,\delta\right) = \mathcal{H}\left(s-\delta\right)
\end{equation}
given in (\ref{eq:s-delta}). This scaling is observed along the diagonal of the tenfold periodic classification as displayed by colours in Table \ref{tab:10-fold_way}.

We now apply these results to Dirac symbols.
Note that by design, if $\mathcal{H}_{c}$ is a Dirac symbol, i.e., it is build from anti-commuting $\boldsymbol{\gamma}$ matrices, then 
$C$ abides $\left\{ \gamma_{i},C\right\} =0$ and $C^{2}=1$ so
$C$ is a $\gamma$ matrix, and $\mathcal{H}_{nc}$ is also a Dirac symbol.
If $\mathcal{H}_{nc}$ is a Dirac symbol, $\gamma_{i}\otimes\sigma_{z}$
are also $\gamma$ matrices anti-commuting with each other and  
with $\sigma_{a}$, so that $\mathcal{H}_{c}$ is a Dirac symbol. These results provide a systematic way to build higher dimension Dirac
symbols for different $s$ symmetry classes. Note that if $\theta$ is a momentum
(position) coordinate then the second term is a $\gamma_a$ $\left(\gamma_s\right)$
matrix. The simplest Dirac symbol is $\mathcal{H}_{0}=h_{0}\left(\boldsymbol{k},\boldsymbol{r}\right)\gamma_{s}^0$ where $\gamma_{s}^0= {\unit}$ for which $p=q=0$. It obviously possess time reversal symmetry but no particle hole symmetry so it is
in $AI$ class i.e. with $s=0$. Using the mappings (\ref{eq:K_result}), it appears that higher dimension Dirac
symbols can be obtained for all real classes as given by $s=p-q\;\text{mod\,8}$. For example for $p=s,q=0$ up to $s=4$ \cite{Teo2010}.

%---- References ------%

% \bibliography{Thesis.bib} 
\input{main.bbl}

    % \ifarXiv
    %     \foreach \x in {1,...,\numbersupplementpages}
    %     {
    %         \clearpage
    %         \includepdf[pages={\x,{}}]{\supplementfilename.pdf}
    %     }
    % \fi

\end{document}

%% file: main.bbl

%% file: main.bbl
\begin{thebibliography}{48}%
\makeatletter
\providecommand \@ifxundefined [1]{%
 \@ifx{#1\undefined}
}%
\providecommand \@ifnum [1]{%
 \ifnum #1\expandafter \@firstoftwo
 \else \expandafter \@secondoftwo
 \fi
}%
\providecommand \@ifx [1]{%
 \ifx #1\expandafter \@firstoftwo
 \else \expandafter \@secondoftwo
 \fi
}%
\providecommand \natexlab [1]{#1}%
\providecommand \enquote  [1]{``#1''}%
\providecommand \bibnamefont  [1]{#1}%
\providecommand \bibfnamefont [1]{#1}%
\providecommand \citenamefont [1]{#1}%
\providecommand \href@noop [0]{\@secondoftwo}%
\providecommand \href [0]{\begingroup \@sanitize@url \@href}%
\providecommand \@href[1]{\@@startlink{#1}\@@href}%
\providecommand \@@href[1]{\endgroup#1\@@endlink}%
\providecommand \@sanitize@url [0]{\catcode `\\12\catcode `\$12\catcode
  `\&12\catcode `\#12\catcode `\^12\catcode `\_12\catcode `\%12\relax}%
\providecommand \@@startlink[1]{}%
\providecommand \@@endlink[0]{}%
\providecommand \url  [0]{\begingroup\@sanitize@url \@url }%
\providecommand \@url [1]{\endgroup\@href {#1}{\urlprefix }}%
\providecommand \urlprefix  [0]{URL }%
\providecommand \Eprint [0]{\href }%
\providecommand \doibase [0]{https://doi.org/}%
\providecommand \selectlanguage [0]{\@gobble}%
\providecommand \bibinfo  [0]{\@secondoftwo}%
\providecommand \bibfield  [0]{\@secondoftwo}%
\providecommand \translation [1]{[#1]}%
\providecommand \BibitemOpen [0]{}%
\providecommand \bibitemStop [0]{}%
\providecommand \bibitemNoStop [0]{.\EOS\space}%
\providecommand \EOS [0]{\spacefactor3000\relax}%
\providecommand \BibitemShut  [1]{\csname bibitem#1\endcsname}%
\let\auto@bib@innerbib\@empty
%</preamble>
\bibitem [{\citenamefont {Altland}\ and\ \citenamefont
  {Zirnbauer}(1997)}]{Altland1997}%
  \BibitemOpen
  \bibfield  {author} {\bibinfo {author} {\bibfnamefont {A.}~\bibnamefont
  {Altland}}\ and\ \bibinfo {author} {\bibfnamefont {M.~R.}\ \bibnamefont
  {Zirnbauer}},\ }\href {https://doi.org/10.1103/PhysRevB.55.1142} {\bibfield
  {journal} {\bibinfo  {journal} {Phys. Rev. B}\ }\textbf {\bibinfo {volume}
  {55}},\ \bibinfo {pages} {1142} (\bibinfo {year} {1997})}\BibitemShut
  {NoStop}%
\bibitem [{\citenamefont {Kitaev}(2009)}]{Kitaev2009}%
  \BibitemOpen
  \bibfield  {author} {\bibinfo {author} {\bibfnamefont {A.}~\bibnamefont
  {Kitaev}},\ }\href {https://doi.org/10.1063/1.3149495} {\bibfield  {journal}
  {\bibinfo  {journal} {AIP Conference Proceedings}\ }\textbf {\bibinfo
  {volume} {1134}},\ \bibinfo {pages} {22} (\bibinfo {year}
  {2009})}\BibitemShut {NoStop}%
\bibitem [{\citenamefont {Schnyder}\ \emph {et~al.}(2008)\citenamefont
  {Schnyder}, \citenamefont {Ryu}, \citenamefont {Furusaki},\ and\
  \citenamefont {Ludwig}}]{Schnyder2008}%
  \BibitemOpen
  \bibfield  {author} {\bibinfo {author} {\bibfnamefont {A.~P.}\ \bibnamefont
  {Schnyder}}, \bibinfo {author} {\bibfnamefont {S.}~\bibnamefont {Ryu}},
  \bibinfo {author} {\bibfnamefont {A.}~\bibnamefont {Furusaki}},\ and\
  \bibinfo {author} {\bibfnamefont {A.~W.~W.}\ \bibnamefont {Ludwig}},\ }\href
  {https://doi.org/10.1103/PhysRevB.78.195125} {\bibfield  {journal} {\bibinfo
  {journal} {Phys. Rev. B}\ }\textbf {\bibinfo {volume} {78}},\ \bibinfo
  {pages} {195125} (\bibinfo {year} {2008})}\BibitemShut {NoStop}%
\bibitem [{\citenamefont {Stone}\ \emph {et~al.}(2010)\citenamefont {Stone},
  \citenamefont {Chiu},\ and\ \citenamefont {Roy}}]{Stone2010}%
  \BibitemOpen
  \bibfield  {author} {\bibinfo {author} {\bibfnamefont {M.}~\bibnamefont
  {Stone}}, \bibinfo {author} {\bibfnamefont {C.-K.}\ \bibnamefont {Chiu}},\
  and\ \bibinfo {author} {\bibfnamefont {A.}~\bibnamefont {Roy}},\ }\href
  {https://doi.org/10.1088/1751-8113/44/4/045001} {\bibfield  {journal}
  {\bibinfo  {journal} {Journal of Physics A: Mathematical and Theoretical}\
  }\textbf {\bibinfo {volume} {44}},\ \bibinfo {pages} {045001} (\bibinfo
  {year} {2010})}\BibitemShut {NoStop}%
\bibitem [{\citenamefont {Teo}\ and\ \citenamefont {Kane}(2010)}]{Teo2010}%
  \BibitemOpen
  \bibfield  {author} {\bibinfo {author} {\bibfnamefont {J.~C.~Y.}\
  \bibnamefont {Teo}}\ and\ \bibinfo {author} {\bibfnamefont {C.~L.}\
  \bibnamefont {Kane}},\ }\href {https://doi.org/10.1103/PhysRevB.82.115120}
  {\bibfield  {journal} {\bibinfo  {journal} {Phys. Rev. B}\ }\textbf {\bibinfo
  {volume} {82}},\ \bibinfo {pages} {115120} (\bibinfo {year}
  {2010})}\BibitemShut {NoStop}%
\bibitem [{\citenamefont {{Toulouse, G.}}\ and\ \citenamefont {{Kl\'eman,
  M.}}(1976)}]{Toulouse1976}%
  \BibitemOpen
  \bibfield  {author} {\bibinfo {author} {\bibnamefont {{Toulouse, G.}}}\ and\
  \bibinfo {author} {\bibnamefont {{Kl\'eman, M.}}},\ }\href
  {https://doi.org/10.1051/jphyslet:01976003706014900} {\bibfield  {journal}
  {\bibinfo  {journal} {J. Physique Lett.}\ }\textbf {\bibinfo {volume} {37}},\
  \bibinfo {pages} {149} (\bibinfo {year} {1976})}\BibitemShut {NoStop}%
\bibitem [{\citenamefont {Kohmoto}(1985)}]{KOHMOTO1985}%
  \BibitemOpen
  \bibfield  {author} {\bibinfo {author} {\bibfnamefont {M.}~\bibnamefont
  {Kohmoto}},\ }\href
  {https://doi.org/https://doi.org/10.1016/0003-4916(85)90148-4} {\bibfield
  {journal} {\bibinfo  {journal} {Annals of Physics}\ }\textbf {\bibinfo
  {volume} {160}},\ \bibinfo {pages} {343} (\bibinfo {year}
  {1985})}\BibitemShut {NoStop}%
\bibitem [{\citenamefont {Thouless}\ \emph {et~al.}(1982)\citenamefont
  {Thouless}, \citenamefont {Kohmoto}, \citenamefont {Nightingale},\ and\
  \citenamefont {den Nijs}}]{TKNN1982}%
  \BibitemOpen
  \bibfield  {author} {\bibinfo {author} {\bibfnamefont {D.~J.}\ \bibnamefont
  {Thouless}}, \bibinfo {author} {\bibfnamefont {M.}~\bibnamefont {Kohmoto}},
  \bibinfo {author} {\bibfnamefont {M.~P.}\ \bibnamefont {Nightingale}},\ and\
  \bibinfo {author} {\bibfnamefont {M.}~\bibnamefont {den Nijs}},\ }\href
  {https://doi.org/10.1103/PhysRevLett.49.405} {\bibfield  {journal} {\bibinfo
  {journal} {Phys. Rev. Lett.}\ }\textbf {\bibinfo {volume} {49}},\ \bibinfo
  {pages} {405} (\bibinfo {year} {1982})}\BibitemShut {NoStop}%
\bibitem [{\citenamefont {Atiyah}\ and\ \citenamefont
  {Singer}(1963)}]{Atiyah1963}%
  \BibitemOpen
  \bibfield  {author} {\bibinfo {author} {\bibfnamefont {M.~F.}\ \bibnamefont
  {Atiyah}}\ and\ \bibinfo {author} {\bibfnamefont {I.~M.}\ \bibnamefont
  {Singer}},\ }\href
  {https://www.ams.org/journals/bull/1963-69-03/S0002-9904-1963-10957-X/}
  {\bibfield  {journal} {\bibinfo  {journal} {Bulletin of the American
  Mathematical Society}\ }\textbf {\bibinfo {volume} {69}},\ \bibinfo {pages}
  {422 } (\bibinfo {year} {1963})}\BibitemShut {NoStop}%
\bibitem [{\citenamefont {Atiyah}\ and\ \citenamefont
  {Singer}(1968)}]{Atiyah1968}%
  \BibitemOpen
  \bibfield  {author} {\bibinfo {author} {\bibfnamefont {M.~F.}\ \bibnamefont
  {Atiyah}}\ and\ \bibinfo {author} {\bibfnamefont {I.~M.}\ \bibnamefont
  {Singer}},\ }\href {http://www.jstor.org/stable/1970715} {\bibfield
  {journal} {\bibinfo  {journal} {Annals of Mathematics}\ }\textbf {\bibinfo
  {volume} {87}},\ \bibinfo {pages} {484} (\bibinfo {year} {1968})}\BibitemShut
  {NoStop}%
\bibitem [{\citenamefont {Nakahara}(1990)}]{Nakahara1990}%
  \BibitemOpen
  \bibfield  {author} {\bibinfo {author} {\bibfnamefont {M.}~\bibnamefont
  {Nakahara}},\ }\href {https://cds.cern.ch/record/206619} {\emph {\bibinfo
  {title} {{Geometry, topology and physics}}}},\ Graduate student series in
  physics\ (\bibinfo  {publisher} {Hilger},\ \bibinfo {address} {Bristol},\
  \bibinfo {year} {1990})\BibitemShut {NoStop}%
\bibitem [{\citenamefont {Stone}(1984)}]{Stone1984}%
  \BibitemOpen
  \bibfield  {author} {\bibinfo {author} {\bibfnamefont {M.}~\bibnamefont
  {Stone}},\ }\href
  {https://doi.org/https://doi.org/10.1016/0003-4916(84)90252-5} {\bibfield
  {journal} {\bibinfo  {journal} {Annals of Physics}\ }\textbf {\bibinfo
  {volume} {155}},\ \bibinfo {pages} {56} (\bibinfo {year} {1984})}\BibitemShut
  {NoStop}%
\bibitem [{\citenamefont {Chiu}\ \emph {et~al.}(2016)\citenamefont {Chiu},
  \citenamefont {Teo}, \citenamefont {Schnyder},\ and\ \citenamefont
  {Ryu}}]{Chiu2016}%
  \BibitemOpen
  \bibfield  {author} {\bibinfo {author} {\bibfnamefont {C.-K.}\ \bibnamefont
  {Chiu}}, \bibinfo {author} {\bibfnamefont {J.~C.~Y.}\ \bibnamefont {Teo}},
  \bibinfo {author} {\bibfnamefont {A.~P.}\ \bibnamefont {Schnyder}},\ and\
  \bibinfo {author} {\bibfnamefont {S.}~\bibnamefont {Ryu}},\ }\href
  {https://doi.org/10.1103/RevModPhys.88.035005} {\bibfield  {journal}
  {\bibinfo  {journal} {Rev. Mod. Phys.}\ }\textbf {\bibinfo {volume} {88}},\
  \bibinfo {pages} {035005} (\bibinfo {year} {2016})}\BibitemShut {NoStop}%
\bibitem [{\citenamefont {Callias}(1978)}]{Callias1978}%
  \BibitemOpen
  \bibfield  {author} {\bibinfo {author} {\bibfnamefont {C.}~\bibnamefont
  {Callias}},\ }\href {https://doi.org/cmp/1103904395} {\bibfield  {journal}
  {\bibinfo  {journal} {Communications in Mathematical Physics}\ }\textbf
  {\bibinfo {volume} {62}},\ \bibinfo {pages} {213 } (\bibinfo {year}
  {1978})}\BibitemShut {NoStop}%
\bibitem [{\citenamefont {Ümit Ertem}(2017)}]{Ertem2017}%
  \BibitemOpen
  \bibfield  {author} {\bibinfo {author} {\bibnamefont {Ümit Ertem}},\ }\href
  {https://doi.org/10.1088/2399-6528/aa8ab7} {\bibfield  {journal} {\bibinfo
  {journal} {Journal of Physics Communications}\ }\textbf {\bibinfo {volume}
  {1}},\ \bibinfo {pages} {035001} (\bibinfo {year} {2017})}\BibitemShut
  {NoStop}%
\bibitem [{\citenamefont {Fukui}\ \emph {et~al.}(2012)\citenamefont {Fukui},
  \citenamefont {Shiozaki}, \citenamefont {Fujiwara},\ and\ \citenamefont
  {Fujimoto}}]{Fukui2012}%
  \BibitemOpen
  \bibfield  {author} {\bibinfo {author} {\bibfnamefont {T.}~\bibnamefont
  {Fukui}}, \bibinfo {author} {\bibfnamefont {K.}~\bibnamefont {Shiozaki}},
  \bibinfo {author} {\bibfnamefont {T.}~\bibnamefont {Fujiwara}},\ and\
  \bibinfo {author} {\bibfnamefont {S.}~\bibnamefont {Fujimoto}},\ }\href
  {https://doi.org/10.1143/JPSJ.81.114602} {\bibfield  {journal} {\bibinfo
  {journal} {Journal of the Physical Society of Japan}\ }\textbf {\bibinfo
  {volume} {81}},\ \bibinfo {pages} {114602} (\bibinfo {year}
  {2012})}\BibitemShut {NoStop}%
\bibitem [{\citenamefont {Eguchi}\ \emph {et~al.}(1980)\citenamefont {Eguchi},
  \citenamefont {Gilkey},\ and\ \citenamefont {Hanson}}]{Eguchi1980}%
  \BibitemOpen
  \bibfield  {author} {\bibinfo {author} {\bibfnamefont {T.}~\bibnamefont
  {Eguchi}}, \bibinfo {author} {\bibfnamefont {P.~B.}\ \bibnamefont {Gilkey}},\
  and\ \bibinfo {author} {\bibfnamefont {A.~J.}\ \bibnamefont {Hanson}},\
  }\href {https://doi.org/10.1016/0370-1573(80)90130-1} {\bibfield  {journal}
  {\bibinfo  {journal} {Phys. Rept.}\ }\textbf {\bibinfo {volume} {66}},\
  \bibinfo {pages} {213} (\bibinfo {year} {1980})}\BibitemShut {NoStop}%
\bibitem [{\citenamefont {Niemi}\ and\ \citenamefont
  {Semenoff}(1984)}]{Niemi&Semenoff1984}%
  \BibitemOpen
  \bibfield  {author} {\bibinfo {author} {\bibfnamefont {A.~J.}\ \bibnamefont
  {Niemi}}\ and\ \bibinfo {author} {\bibfnamefont {G.~W.}\ \bibnamefont
  {Semenoff}},\ }\href {https://doi.org/10.1103/PhysRevD.30.809} {\bibfield
  {journal} {\bibinfo  {journal} {Phys. Rev. D}\ }\textbf {\bibinfo {volume}
  {30}},\ \bibinfo {pages} {809} (\bibinfo {year} {1984})}\BibitemShut
  {NoStop}%
\bibitem [{\citenamefont {Getzler}(1983)}]{Getzler1983}%
  \BibitemOpen
  \bibfield  {author} {\bibinfo {author} {\bibfnamefont {E.}~\bibnamefont
  {Getzler}},\ }\href {https://doi.org/10.1007/BF01210843} {\bibfield
  {journal} {\bibinfo  {journal} {Communications in Mathematical Physics}\
  }\textbf {\bibinfo {volume} {92}},\ \bibinfo {pages} {163 } (\bibinfo {year}
  {1983})}\BibitemShut {NoStop}%
\bibitem [{\citenamefont {{Akkermans, E.}}\ \emph {et~al.}(1998)\citenamefont
  {{Akkermans, E.}}, \citenamefont {{Avron, J. E.}}, \citenamefont {{Narevich,
  R.}},\ and\ \citenamefont {{Seiler, R.}}}]{Akkermans1998}%
  \BibitemOpen
  \bibfield  {author} {\bibinfo {author} {\bibnamefont {{Akkermans, E.}}},
  \bibinfo {author} {\bibnamefont {{Avron, J. E.}}}, \bibinfo {author}
  {\bibnamefont {{Narevich, R.}}},\ and\ \bibinfo {author} {\bibnamefont
  {{Seiler, R.}}},\ }\href {https://doi.org/10.1007/s100510050160} {\bibfield
  {journal} {\bibinfo  {journal} {Eur. Phys. J. B}\ }\textbf {\bibinfo {volume}
  {1}},\ \bibinfo {pages} {117} (\bibinfo {year} {1998})}\BibitemShut {NoStop}%
\bibitem [{\citenamefont {Hillery}\ \emph {et~al.}(1984)\citenamefont
  {Hillery}, \citenamefont {O'Connell}, \citenamefont {Scully},\ and\
  \citenamefont {Wigner}}]{Hillery1984}%
  \BibitemOpen
  \bibfield  {author} {\bibinfo {author} {\bibfnamefont {M.}~\bibnamefont
  {Hillery}}, \bibinfo {author} {\bibfnamefont {R.}~\bibnamefont {O'Connell}},
  \bibinfo {author} {\bibfnamefont {M.}~\bibnamefont {Scully}},\ and\ \bibinfo
  {author} {\bibfnamefont {E.}~\bibnamefont {Wigner}},\ }\href
  {https://doi.org/https://doi.org/10.1016/0370-1573(84)90160-1} {\bibfield
  {journal} {\bibinfo  {journal} {Physics Reports}\ }\textbf {\bibinfo {volume}
  {106}},\ \bibinfo {pages} {121} (\bibinfo {year} {1984})}\BibitemShut
  {NoStop}%
\bibitem [{\citenamefont {Case}(2008)}]{Case2008}%
  \BibitemOpen
  \bibfield  {author} {\bibinfo {author} {\bibfnamefont {W.~B.}\ \bibnamefont
  {Case}},\ }\href {https://doi.org/10.1119/1.2957889} {\bibfield  {journal}
  {\bibinfo  {journal} {American Journal of Physics}\ }\textbf {\bibinfo
  {volume} {76}},\ \bibinfo {pages} {937} (\bibinfo {year} {2008})}\BibitemShut
  {NoStop}%
\bibitem [{\citenamefont {Yankowsky}\ \emph {et~al.}(2013)\citenamefont
  {Yankowsky}, \citenamefont {Schwarz},\ and\ \citenamefont
  {Levy}}]{yankowsky2013}%
  \BibitemOpen
  \bibfield  {author} {\bibinfo {author} {\bibfnamefont {E.}~\bibnamefont
  {Yankowsky}}, \bibinfo {author} {\bibfnamefont {A.}~\bibnamefont {Schwarz}},\
  and\ \bibinfo {author} {\bibfnamefont {S.}~\bibnamefont {Levy}},\ }\href
  {https://books.google.co.il/books?id=RhXpCAAAQBAJ} {\emph {\bibinfo {title}
  {Quantum Field Theory and Topology}}},\ Grundlehren der mathematischen
  Wissenschaften\ (\bibinfo  {publisher} {Springer Berlin Heidelberg},\
  \bibinfo {year} {2013})\BibitemShut {NoStop}%
\bibitem [{\citenamefont {Castro~Neto}\ \emph {et~al.}(2009)\citenamefont
  {Castro~Neto}, \citenamefont {Guinea}, \citenamefont {Peres}, \citenamefont
  {Novoselov},\ and\ \citenamefont {Geim}}]{Neto2009}%
  \BibitemOpen
  \bibfield  {author} {\bibinfo {author} {\bibfnamefont {A.~H.}\ \bibnamefont
  {Castro~Neto}}, \bibinfo {author} {\bibfnamefont {F.}~\bibnamefont {Guinea}},
  \bibinfo {author} {\bibfnamefont {N.~M.~R.}\ \bibnamefont {Peres}}, \bibinfo
  {author} {\bibfnamefont {K.~S.}\ \bibnamefont {Novoselov}},\ and\ \bibinfo
  {author} {\bibfnamefont {A.~K.}\ \bibnamefont {Geim}},\ }\href
  {https://doi.org/10.1103/RevModPhys.81.109} {\bibfield  {journal} {\bibinfo
  {journal} {Rev. Mod. Phys.}\ }\textbf {\bibinfo {volume} {81}},\ \bibinfo
  {pages} {109} (\bibinfo {year} {2009})}\BibitemShut {NoStop}%
\bibitem [{\citenamefont {Kelly}\ and\ \citenamefont
  {Halas}(1998)}]{Kelly1998}%
  \BibitemOpen
  \bibfield  {author} {\bibinfo {author} {\bibfnamefont {K.}~\bibnamefont
  {Kelly}}\ and\ \bibinfo {author} {\bibfnamefont {N.}~\bibnamefont {Halas}},\
  }\href {https://doi.org/https://doi.org/10.1016/S0039-6028(98)00622-0}
  {\bibfield  {journal} {\bibinfo  {journal} {Surface Science}\ }\textbf
  {\bibinfo {volume} {416}},\ \bibinfo {pages} {L1085} (\bibinfo {year}
  {1998})}\BibitemShut {NoStop}%
\bibitem [{\citenamefont {Ugeda}\ \emph {et~al.}(2010)\citenamefont {Ugeda},
  \citenamefont {Brihuega}, \citenamefont {Guinea},\ and\ \citenamefont
  {G\'omez-Rodr\'iguez}}]{Ugeda2010}%
  \BibitemOpen
  \bibfield  {author} {\bibinfo {author} {\bibfnamefont {M.~M.}\ \bibnamefont
  {Ugeda}}, \bibinfo {author} {\bibfnamefont {I.}~\bibnamefont {Brihuega}},
  \bibinfo {author} {\bibfnamefont {F.}~\bibnamefont {Guinea}},\ and\ \bibinfo
  {author} {\bibfnamefont {J.~M.}\ \bibnamefont {G\'omez-Rodr\'iguez}},\ }\href
  {https://doi.org/10.1103/PhysRevLett.104.096804} {\bibfield  {journal}
  {\bibinfo  {journal} {Phys. Rev. Lett.}\ }\textbf {\bibinfo {volume} {104}},\
  \bibinfo {pages} {096804} (\bibinfo {year} {2010})}\BibitemShut {NoStop}%
\bibitem [{\citenamefont {{Ovdat}}\ \emph {et~al.}(2017)\citenamefont
  {{Ovdat}}, \citenamefont {{Mao}}, \citenamefont {{Jiang}}, \citenamefont
  {{Andrei}},\ and\ \citenamefont {{Akkermans}}}]{Ovdat2017}%
  \BibitemOpen
  \bibfield  {author} {\bibinfo {author} {\bibfnamefont {O.}~\bibnamefont
  {{Ovdat}}}, \bibinfo {author} {\bibfnamefont {J.}~\bibnamefont {{Mao}}},
  \bibinfo {author} {\bibfnamefont {Y.}~\bibnamefont {{Jiang}}}, \bibinfo
  {author} {\bibfnamefont {E.~Y.}\ \bibnamefont {{Andrei}}},\ and\ \bibinfo
  {author} {\bibfnamefont {E.}~\bibnamefont {{Akkermans}}},\ }\href
  {https://doi.org/10.1038/s41467-017-00591-8} {\bibfield  {journal} {\bibinfo
  {journal} {Nature Communications}\ }\textbf {\bibinfo {volume} {8}} (\bibinfo
  {year} {2017})}\BibitemShut {NoStop}%
\bibitem [{\citenamefont {Pereira}\ \emph {et~al.}(2006)\citenamefont
  {Pereira}, \citenamefont {Guinea}, \citenamefont {Lopes~dos Santos},
  \citenamefont {Peres},\ and\ \citenamefont {Castro~Neto}}]{Pereira2006}%
  \BibitemOpen
  \bibfield  {author} {\bibinfo {author} {\bibfnamefont {V.~M.}\ \bibnamefont
  {Pereira}}, \bibinfo {author} {\bibfnamefont {F.}~\bibnamefont {Guinea}},
  \bibinfo {author} {\bibfnamefont {J.~M.~B.}\ \bibnamefont {Lopes~dos
  Santos}}, \bibinfo {author} {\bibfnamefont {N.~M.~R.}\ \bibnamefont
  {Peres}},\ and\ \bibinfo {author} {\bibfnamefont {A.~H.}\ \bibnamefont
  {Castro~Neto}},\ }\href {https://doi.org/10.1103/PhysRevLett.96.036801}
  {\bibfield  {journal} {\bibinfo  {journal} {Phys. Rev. Lett.}\ }\textbf
  {\bibinfo {volume} {96}},\ \bibinfo {pages} {036801} (\bibinfo {year}
  {2006})}\BibitemShut {NoStop}%
\bibitem [{\citenamefont {Pereira}\ \emph {et~al.}(2008)\citenamefont
  {Pereira}, \citenamefont {Lopes~dos Santos},\ and\ \citenamefont
  {Castro~Neto}}]{Pereira2008}%
  \BibitemOpen
  \bibfield  {author} {\bibinfo {author} {\bibfnamefont {V.~M.}\ \bibnamefont
  {Pereira}}, \bibinfo {author} {\bibfnamefont {J.~M.~B.}\ \bibnamefont
  {Lopes~dos Santos}},\ and\ \bibinfo {author} {\bibfnamefont {A.~H.}\
  \bibnamefont {Castro~Neto}},\ }\href
  {https://doi.org/10.1103/PhysRevB.77.115109} {\bibfield  {journal} {\bibinfo
  {journal} {Phys. Rev. B}\ }\textbf {\bibinfo {volume} {77}},\ \bibinfo
  {pages} {115109} (\bibinfo {year} {2008})}\BibitemShut {NoStop}%
\bibitem [{\citenamefont {Amara}\ \emph {et~al.}(2007)\citenamefont {Amara},
  \citenamefont {Latil}, \citenamefont {Meunier}, \citenamefont {Lambin},\ and\
  \citenamefont {Charlier}}]{Amara2007}%
  \BibitemOpen
  \bibfield  {author} {\bibinfo {author} {\bibfnamefont {H.}~\bibnamefont
  {Amara}}, \bibinfo {author} {\bibfnamefont {S.}~\bibnamefont {Latil}},
  \bibinfo {author} {\bibfnamefont {V.}~\bibnamefont {Meunier}}, \bibinfo
  {author} {\bibfnamefont {P.}~\bibnamefont {Lambin}},\ and\ \bibinfo {author}
  {\bibfnamefont {J.-C.}\ \bibnamefont {Charlier}},\ }\href
  {https://doi.org/10.1103/PhysRevB.76.115423} {\bibfield  {journal} {\bibinfo
  {journal} {Phys. Rev. B}\ }\textbf {\bibinfo {volume} {76}},\ \bibinfo
  {pages} {115423} (\bibinfo {year} {2007})}\BibitemShut {NoStop}%
\bibitem [{\citenamefont {Palacios}\ \emph {et~al.}(2008)\citenamefont
  {Palacios}, \citenamefont {Fern\'andez-Rossier},\ and\ \citenamefont
  {Brey}}]{Palacios2008}%
  \BibitemOpen
  \bibfield  {author} {\bibinfo {author} {\bibfnamefont {J.~J.}\ \bibnamefont
  {Palacios}}, \bibinfo {author} {\bibfnamefont {J.}~\bibnamefont
  {Fern\'andez-Rossier}},\ and\ \bibinfo {author} {\bibfnamefont
  {L.}~\bibnamefont {Brey}},\ }\href
  {https://doi.org/10.1103/PhysRevB.77.195428} {\bibfield  {journal} {\bibinfo
  {journal} {Phys. Rev. B}\ }\textbf {\bibinfo {volume} {77}},\ \bibinfo
  {pages} {195428} (\bibinfo {year} {2008})}\BibitemShut {NoStop}%
\bibitem [{\citenamefont {Dutreix}\ \emph {et~al.}(2013)\citenamefont
  {Dutreix}, \citenamefont {Bilteanu}, \citenamefont {Jagannathan},\ and\
  \citenamefont {Bena}}]{Dutreix2013}%
  \BibitemOpen
  \bibfield  {author} {\bibinfo {author} {\bibfnamefont {C.}~\bibnamefont
  {Dutreix}}, \bibinfo {author} {\bibfnamefont {L.}~\bibnamefont {Bilteanu}},
  \bibinfo {author} {\bibfnamefont {A.}~\bibnamefont {Jagannathan}},\ and\
  \bibinfo {author} {\bibfnamefont {C.}~\bibnamefont {Bena}},\ }\href
  {https://doi.org/10.1103/PhysRevB.87.245413} {\bibfield  {journal} {\bibinfo
  {journal} {Phys. Rev. B}\ }\textbf {\bibinfo {volume} {87}},\ \bibinfo
  {pages} {245413} (\bibinfo {year} {2013})}\BibitemShut {NoStop}%
\bibitem [{\citenamefont {Mallet}\ \emph {et~al.}(2012)\citenamefont {Mallet},
  \citenamefont {Brihuega}, \citenamefont {Bose}, \citenamefont {Ugeda},
  \citenamefont {G\'omez-Rodr\'{\i}guez}, \citenamefont {Kern},\ and\
  \citenamefont {Veuillen}}]{Mallet2012}%
  \BibitemOpen
  \bibfield  {author} {\bibinfo {author} {\bibfnamefont {P.}~\bibnamefont
  {Mallet}}, \bibinfo {author} {\bibfnamefont {I.}~\bibnamefont {Brihuega}},
  \bibinfo {author} {\bibfnamefont {S.}~\bibnamefont {Bose}}, \bibinfo {author}
  {\bibfnamefont {M.~M.}\ \bibnamefont {Ugeda}}, \bibinfo {author}
  {\bibfnamefont {J.~M.}\ \bibnamefont {G\'omez-Rodr\'{\i}guez}}, \bibinfo
  {author} {\bibfnamefont {K.}~\bibnamefont {Kern}},\ and\ \bibinfo {author}
  {\bibfnamefont {J.~Y.}\ \bibnamefont {Veuillen}},\ }\href
  {https://doi.org/10.1103/PhysRevB.86.045444} {\bibfield  {journal} {\bibinfo
  {journal} {Phys. Rev. B}\ }\textbf {\bibinfo {volume} {86}},\ \bibinfo
  {pages} {045444} (\bibinfo {year} {2012})}\BibitemShut {NoStop}%
\bibitem [{\citenamefont {Ovdat}\ \emph {et~al.}(2020)\citenamefont {Ovdat},
  \citenamefont {Don},\ and\ \citenamefont {Akkermans}}]{Ovdat2020}%
  \BibitemOpen
  \bibfield  {author} {\bibinfo {author} {\bibfnamefont {O.}~\bibnamefont
  {Ovdat}}, \bibinfo {author} {\bibfnamefont {Y.}~\bibnamefont {Don}},\ and\
  \bibinfo {author} {\bibfnamefont {E.}~\bibnamefont {Akkermans}},\ }\href
  {https://doi.org/10.1103/PhysRevB.102.075109} {\bibfield  {journal} {\bibinfo
   {journal} {Phys. Rev. B}\ }\textbf {\bibinfo {volume} {102}},\ \bibinfo
  {pages} {075109} (\bibinfo {year} {2020})}\BibitemShut {NoStop}%
\bibitem [{\citenamefont {Su}\ \emph {et~al.}(1979)\citenamefont {Su},
  \citenamefont {Schrieffer},\ and\ \citenamefont {Heeger}}]{SSH1979}%
  \BibitemOpen
  \bibfield  {author} {\bibinfo {author} {\bibfnamefont {W.~P.}\ \bibnamefont
  {Su}}, \bibinfo {author} {\bibfnamefont {J.~R.}\ \bibnamefont {Schrieffer}},\
  and\ \bibinfo {author} {\bibfnamefont {A.~J.}\ \bibnamefont {Heeger}},\
  }\href {https://doi.org/10.1103/PhysRevLett.42.1698} {\bibfield  {journal}
  {\bibinfo  {journal} {Phys. Rev. Lett.}\ }\textbf {\bibinfo {volume} {42}},\
  \bibinfo {pages} {1698} (\bibinfo {year} {1979})}\BibitemShut {NoStop}%
\bibitem [{\citenamefont {Heeger}\ \emph {et~al.}(1988)\citenamefont {Heeger},
  \citenamefont {Kivelson}, \citenamefont {Schrieffer},\ and\ \citenamefont
  {Su}}]{Heeger1988}%
  \BibitemOpen
  \bibfield  {author} {\bibinfo {author} {\bibfnamefont {A.~J.}\ \bibnamefont
  {Heeger}}, \bibinfo {author} {\bibfnamefont {S.}~\bibnamefont {Kivelson}},
  \bibinfo {author} {\bibfnamefont {J.~R.}\ \bibnamefont {Schrieffer}},\ and\
  \bibinfo {author} {\bibfnamefont {W.~P.}\ \bibnamefont {Su}},\ }\href
  {https://doi.org/10.1103/RevModPhys.60.781} {\bibfield  {journal} {\bibinfo
  {journal} {Rev. Mod. Phys.}\ }\textbf {\bibinfo {volume} {60}},\ \bibinfo
  {pages} {781} (\bibinfo {year} {1988})}\BibitemShut {NoStop}%
\bibitem [{\citenamefont {Dutreix}\ and\ \citenamefont
  {Katsnelson}(2016)}]{Dutreix2016}%
  \BibitemOpen
  \bibfield  {author} {\bibinfo {author} {\bibfnamefont {C.}~\bibnamefont
  {Dutreix}}\ and\ \bibinfo {author} {\bibfnamefont {M.~I.}\ \bibnamefont
  {Katsnelson}},\ }\href {https://doi.org/10.1103/PhysRevB.93.035413}
  {\bibfield  {journal} {\bibinfo  {journal} {Phys. Rev. B}\ }\textbf {\bibinfo
  {volume} {93}},\ \bibinfo {pages} {035413} (\bibinfo {year}
  {2016})}\BibitemShut {NoStop}%
\bibitem [{\citenamefont {Nanda}\ \emph {et~al.}(2013)\citenamefont {Nanda},
  \citenamefont {Sherafati}, \citenamefont {Popović},\ and\ \citenamefont
  {Satpathy}}]{Nanda2013}%
  \BibitemOpen
  \bibfield  {author} {\bibinfo {author} {\bibfnamefont {B.~R.~K.}\
  \bibnamefont {Nanda}}, \bibinfo {author} {\bibfnamefont {M.}~\bibnamefont
  {Sherafati}}, \bibinfo {author} {\bibfnamefont {Z.~S.}\ \bibnamefont
  {Popović}},\ and\ \bibinfo {author} {\bibfnamefont {S.}~\bibnamefont
  {Satpathy}},\ }\href {https://doi.org/10.1088/1367-2630/15/3/039501}
  {\bibfield  {journal} {\bibinfo  {journal} {New Journal of Physics}\ }\textbf
  {\bibinfo {volume} {15}},\ \bibinfo {pages} {039501} (\bibinfo {year}
  {2013})}\BibitemShut {NoStop}%
\bibitem [{\citenamefont {Hou}\ \emph {et~al.}(2007)\citenamefont {Hou},
  \citenamefont {Chamon},\ and\ \citenamefont {Mudry}}]{Hou2007}%
  \BibitemOpen
  \bibfield  {author} {\bibinfo {author} {\bibfnamefont {C.-Y.}\ \bibnamefont
  {Hou}}, \bibinfo {author} {\bibfnamefont {C.}~\bibnamefont {Chamon}},\ and\
  \bibinfo {author} {\bibfnamefont {C.}~\bibnamefont {Mudry}},\ }\href
  {https://doi.org/10.1103/PhysRevLett.98.186809} {\bibfield  {journal}
  {\bibinfo  {journal} {Phys. Rev. Lett.}\ }\textbf {\bibinfo {volume} {98}},\
  \bibinfo {pages} {186809} (\bibinfo {year} {2007})}\BibitemShut {NoStop}%
\bibitem [{\citenamefont {Jackiw}\ and\ \citenamefont {Pi}(2007)}]{Jackiw2007}%
  \BibitemOpen
  \bibfield  {author} {\bibinfo {author} {\bibfnamefont {R.}~\bibnamefont
  {Jackiw}}\ and\ \bibinfo {author} {\bibfnamefont {S.-Y.}\ \bibnamefont
  {Pi}},\ }\href {https://doi.org/10.1103/PhysRevLett.98.266402} {\bibfield
  {journal} {\bibinfo  {journal} {Phys. Rev. Lett.}\ }\textbf {\bibinfo
  {volume} {98}},\ \bibinfo {pages} {266402} (\bibinfo {year}
  {2007})}\BibitemShut {NoStop}%
\bibitem [{\citenamefont {Gomes}\ \emph {et~al.}(2012)\citenamefont {Gomes},
  \citenamefont {Mar}, \citenamefont {Ko}, \citenamefont {Guinea},\ and\
  \citenamefont {Manoharan}}]{Gomes2012}%
  \BibitemOpen
  \bibfield  {author} {\bibinfo {author} {\bibfnamefont {K.~K.}\ \bibnamefont
  {Gomes}}, \bibinfo {author} {\bibfnamefont {W.}~\bibnamefont {Mar}}, \bibinfo
  {author} {\bibfnamefont {W.}~\bibnamefont {Ko}}, \bibinfo {author}
  {\bibfnamefont {F.}~\bibnamefont {Guinea}},\ and\ \bibinfo {author}
  {\bibfnamefont {H.~C.}\ \bibnamefont {Manoharan}},\ }\href
  {https://doi.org/10.1038/nature10941} {\bibfield  {journal} {\bibinfo
  {journal} {Nature}\ }\textbf {\bibinfo {volume} {483}},\ \bibinfo {pages}
  {306} (\bibinfo {year} {2012})}\BibitemShut {NoStop}%
\bibitem [{\citenamefont {Qu}\ \emph {et~al.}(2022)\citenamefont {Qu},
  \citenamefont {Nigge}, \citenamefont {Link}, \citenamefont {Levy},
  \citenamefont {Michiardi}, \citenamefont {Spandar}, \citenamefont {Matthé},
  \citenamefont {Schneider}, \citenamefont {Zhdanovich}, \citenamefont
  {Starke}, \citenamefont {Gutiérrez},\ and\ \citenamefont
  {Damascelli}}]{Qu2022}%
  \BibitemOpen
  \bibfield  {author} {\bibinfo {author} {\bibfnamefont {A.~C.}\ \bibnamefont
  {Qu}}, \bibinfo {author} {\bibfnamefont {P.}~\bibnamefont {Nigge}}, \bibinfo
  {author} {\bibfnamefont {S.}~\bibnamefont {Link}}, \bibinfo {author}
  {\bibfnamefont {G.}~\bibnamefont {Levy}}, \bibinfo {author} {\bibfnamefont
  {M.}~\bibnamefont {Michiardi}}, \bibinfo {author} {\bibfnamefont {P.~L.}\
  \bibnamefont {Spandar}}, \bibinfo {author} {\bibfnamefont {T.}~\bibnamefont
  {Matthé}}, \bibinfo {author} {\bibfnamefont {M.}~\bibnamefont {Schneider}},
  \bibinfo {author} {\bibfnamefont {S.}~\bibnamefont {Zhdanovich}}, \bibinfo
  {author} {\bibfnamefont {U.}~\bibnamefont {Starke}}, \bibinfo {author}
  {\bibfnamefont {C.}~\bibnamefont {Gutiérrez}},\ and\ \bibinfo {author}
  {\bibfnamefont {A.}~\bibnamefont {Damascelli}},\ }\href
  {https://doi.org/10.1126/sciadv.abm5180} {\bibfield  {journal} {\bibinfo
  {journal} {Science Advances}\ }\textbf {\bibinfo {volume} {8}},\ \bibinfo
  {pages} {eabm5180} (\bibinfo {year} {2022})}\BibitemShut {NoStop}%
\bibitem [{\citenamefont {Bao}\ \emph {et~al.}(2021)\citenamefont {Bao},
  \citenamefont {Zhang}, \citenamefont {Zhang}, \citenamefont {Wu},
  \citenamefont {Luo}, \citenamefont {Zhou}, \citenamefont {Li}, \citenamefont
  {Hou}, \citenamefont {Yao}, \citenamefont {Liu}, \citenamefont {Yu},
  \citenamefont {Li}, \citenamefont {Duan}, \citenamefont {Yao}, \citenamefont
  {Wang},\ and\ \citenamefont {Zhou}}]{Bao2021}%
  \BibitemOpen
  \bibfield  {author} {\bibinfo {author} {\bibfnamefont {C.}~\bibnamefont
  {Bao}}, \bibinfo {author} {\bibfnamefont {H.}~\bibnamefont {Zhang}}, \bibinfo
  {author} {\bibfnamefont {T.}~\bibnamefont {Zhang}}, \bibinfo {author}
  {\bibfnamefont {X.}~\bibnamefont {Wu}}, \bibinfo {author} {\bibfnamefont
  {L.}~\bibnamefont {Luo}}, \bibinfo {author} {\bibfnamefont {S.}~\bibnamefont
  {Zhou}}, \bibinfo {author} {\bibfnamefont {Q.}~\bibnamefont {Li}}, \bibinfo
  {author} {\bibfnamefont {Y.}~\bibnamefont {Hou}}, \bibinfo {author}
  {\bibfnamefont {W.}~\bibnamefont {Yao}}, \bibinfo {author} {\bibfnamefont
  {L.}~\bibnamefont {Liu}}, \bibinfo {author} {\bibfnamefont {P.}~\bibnamefont
  {Yu}}, \bibinfo {author} {\bibfnamefont {J.}~\bibnamefont {Li}}, \bibinfo
  {author} {\bibfnamefont {W.}~\bibnamefont {Duan}}, \bibinfo {author}
  {\bibfnamefont {H.}~\bibnamefont {Yao}}, \bibinfo {author} {\bibfnamefont
  {Y.}~\bibnamefont {Wang}},\ and\ \bibinfo {author} {\bibfnamefont
  {S.}~\bibnamefont {Zhou}},\ }\href
  {https://doi.org/10.1103/PhysRevLett.126.206804} {\bibfield  {journal}
  {\bibinfo  {journal} {Phys. Rev. Lett.}\ }\textbf {\bibinfo {volume} {126}},\
  \bibinfo {pages} {206804} (\bibinfo {year} {2021})}\BibitemShut {NoStop}%
\bibitem [{\citenamefont {Ryu}\ and\ \citenamefont {Hatsugai}(2002)}]{Ryu2002}%
  \BibitemOpen
  \bibfield  {author} {\bibinfo {author} {\bibfnamefont {S.}~\bibnamefont
  {Ryu}}\ and\ \bibinfo {author} {\bibfnamefont {Y.}~\bibnamefont {Hatsugai}},\
  }\href {https://doi.org/10.1103/PhysRevLett.89.077002} {\bibfield  {journal}
  {\bibinfo  {journal} {Phys. Rev. Lett.}\ }\textbf {\bibinfo {volume} {89}},\
  \bibinfo {pages} {077002} (\bibinfo {year} {2002})}\BibitemShut {NoStop}%
\bibitem [{\citenamefont {Brouwer}\ \emph {et~al.}(2002)\citenamefont
  {Brouwer}, \citenamefont {Racine}, \citenamefont {Furusaki}, \citenamefont
  {Hatsugai}, \citenamefont {Morita},\ and\ \citenamefont
  {Mudry}}]{Brouwer2002}%
  \BibitemOpen
  \bibfield  {author} {\bibinfo {author} {\bibfnamefont {P.~W.}\ \bibnamefont
  {Brouwer}}, \bibinfo {author} {\bibfnamefont {E.}~\bibnamefont {Racine}},
  \bibinfo {author} {\bibfnamefont {A.}~\bibnamefont {Furusaki}}, \bibinfo
  {author} {\bibfnamefont {Y.}~\bibnamefont {Hatsugai}}, \bibinfo {author}
  {\bibfnamefont {Y.}~\bibnamefont {Morita}},\ and\ \bibinfo {author}
  {\bibfnamefont {C.}~\bibnamefont {Mudry}},\ }\href
  {https://doi.org/10.1103/PhysRevB.66.014204} {\bibfield  {journal} {\bibinfo
  {journal} {Phys. Rev. B}\ }\textbf {\bibinfo {volume} {66}},\ \bibinfo
  {pages} {014204} (\bibinfo {year} {2002})}\BibitemShut {NoStop}%
\bibitem [{\citenamefont {Ganeshan}\ \emph {et~al.}(2013)\citenamefont
  {Ganeshan}, \citenamefont {Sun},\ and\ \citenamefont
  {Das~Sarma}}]{Ganeshan2013}%
  \BibitemOpen
  \bibfield  {author} {\bibinfo {author} {\bibfnamefont {S.}~\bibnamefont
  {Ganeshan}}, \bibinfo {author} {\bibfnamefont {K.}~\bibnamefont {Sun}},\ and\
  \bibinfo {author} {\bibfnamefont {S.}~\bibnamefont {Das~Sarma}},\ }\href
  {https://doi.org/10.1103/PhysRevLett.110.180403} {\bibfield  {journal}
  {\bibinfo  {journal} {Phys. Rev. Lett.}\ }\textbf {\bibinfo {volume} {110}},\
  \bibinfo {pages} {180403} (\bibinfo {year} {2013})}\BibitemShut {NoStop}%
\bibitem [{\citenamefont {Lieb}(1989)}]{lieb1989}%
  \BibitemOpen
  \bibfield  {author} {\bibinfo {author} {\bibfnamefont {E.~H.}\ \bibnamefont
  {Lieb}},\ }\href {https://doi.org/10.1103/PhysRevLett.62.1201} {\bibfield
  {journal} {\bibinfo  {journal} {Phys. Rev. Lett.}\ }\textbf {\bibinfo
  {volume} {62}},\ \bibinfo {pages} {1201} (\bibinfo {year}
  {1989})}\BibitemShut {NoStop}%
\bibitem [{\citenamefont {Venaille}\ and\ \citenamefont
  {Delplace}(2021)}]{Venaille2021}%
  \BibitemOpen
  \bibfield  {author} {\bibinfo {author} {\bibfnamefont {A.}~\bibnamefont
  {Venaille}}\ and\ \bibinfo {author} {\bibfnamefont {P.}~\bibnamefont
  {Delplace}},\ }\href {https://doi.org/10.1103/PhysRevResearch.3.043002}
  {\bibfield  {journal} {\bibinfo  {journal} {Phys. Rev. Res.}\ }\textbf
  {\bibinfo {volume} {3}},\ \bibinfo {pages} {043002} (\bibinfo {year}
  {2021})}\BibitemShut {NoStop}%
\end{thebibliography}
